\documentclass[prodmode,acmtecs]{acmsmall} 

\usepackage{multirow}
\usepackage[ruled]{algorithm2e}
\renewcommand{\algorithmcfname}{ALGORITHM}
\SetAlFnt{\small}
\SetAlCapFnt{\small}
\SetAlCapNameFnt{\small}
\SetAlCapHSkip{0pt}
\IncMargin{-\parindent}

\graphicspath{{images/pdf/}}

\begin{document}

\markboth{M. Ramezani, H.R. Rabiee, M. Tahani and A. Rajabi}{DANI: A Fast Community-Preserving, Diffusion Aware Network Inference Algorithm}

\title{DANI: A Fast Diffusion Aware Network Inference Algorithm}
\author{MARYAM RAMEZANI
\affil{Sharif University of Technology}
HAMID R. RABIEE
\affil{Sharif University of Technology}
MARYAM TAHANI
\affil{Sharif University of Technology}
AREZOO RAJABI
\affil{Sharif University of Technology}
}

\begin{abstract}
The fast growth of social networks and their privacy requirements in recent years, has lead to increasing difficulty in obtaining complete topology of these networks. However, diffusion information over these networks is available and many algorithms have been proposed to infer the underlying networks by using this information. The previously proposed algorithms only focus on inferring more links and do not pay attention to the important characteristics of the underlying social networks 
In this paper, we propose a novel algorithm, called DANI, to infer the underlying network structure while preserving its 
properties by using the diffusion information. Moreover, the running time of the proposed method is considerably lower than the previous methods. We applied the proposed method to both real and synthetic networks. The experimental results showed that DANI has higher accuracy and lower run time compared to well-known network inference methods.
\end{abstract}




\acmformat{Maryam Ramezani, Hamid R. Rabiee, Maryam Tahani and Arezoo Rajabi, 2014. DANI: A Fast Community-Preserving, Diffusion Aware Network Inference Algorithm. ACM Transaction on Intelligent Systems and Technology.}

\begin{bottomstuff}

Author's addresses: M. Ramezani (m\underline{ }ramezani@ce.sharif.edu), H. R. Rabiee (rabiee@sharif.edu), M. Tahani (tahani@ce.sharif.edu) and A. Rajabi (arezoorajabi@ce.sharif.edu), Department of Computer Engineering, Sharif University of Technology, Tehran, Iran.

\end{bottomstuff}

\maketitle

\section{Introduction}
\label{sec:introduction}

Online Social Networks (OSNs) that play an important role in the exchange of information between people, have grown noticeably in the last few years. \textit{Diffusion} is a fundamental process over these networks by which information, ideas and new behaviors, namely contagion, disseminate over the network \cite{Netinf:2010}. Propagation of a contagion over a network creates a trace that is called \textit{cascade} (Fig.~\ref{fig:cascade}) \cite{Netinf:2010}. These cascades represent the diffusion behavior that convey valuable information about the underlying network.

Some previous studies indicate that diffusion behavior and network structure are tightly related \cite{easley2010,Eftekhar:2013,BarbieriBM13,DBLP:conf/icdm/BarbieriBM13}.  In other words, actions of users do not only emanate from their individual interests but also depend on the interaction pattern between them. Understanding this relation can result in a more accurate analysis of social networks and the processes taking place over them (e.g. diffusion).

One of the most important features of OSNs is their \textit{modular structure}. Each module in the network is a dense sub-graph called a \textit{community}. In other words, a community is a set of nodes which are densely connected to each other, and sparsely to the rest of nodes. Nodes in a same community have common interests and similar actions \cite{FuzzySystemsChen,fortunato2010}. For example, strongly connected communities of global web networks often have pages with common topics \cite{fortunato2010}. 

In many situations, the underlying network topology (nodes and the links between them) may not be achievable for different reasons: 1)Rapid changes in the structure of networks over time caused by the arrival and departure of members and alteration in their friendships, and 2)Limited access to data due to the policy of OSNs and privacy of user profiles.
Network topology inference is a solution to surmount the above difficulties.

As mentioned before, the diffusion process is influenced by structure of the latent underlying network which includes its community structure.  Thus, observing the diffusion behavior of nodes can help us to gain insight into the networks community structure. In most cases, the only available observation is the time when a contagion reaches a node, namely the infection time of a node \cite{Netinf:2010,nipsMyersL10,EslamiRS11,Netrate11,Fastinf11,journals/abs-1205-1671}. Network inference includes a class of studies, which aim to discover the latent network by utilizing these infection times in different contagions. Several network inference algorithms have been proposed, but most of them do not preserve the community structure during the inference process. Despite their acceptable accuracy on detecting network links, the community structure of the inferred network does not match with the original network. 

Many community-based services are provided on OSNs. Since gathering information about the entire network topology is often impossible, usually inferred networks from the set of observations are being used. By considering community structures during network inference, the community-based services such as optimizing search engines \cite{zhong2008topseer}, visualizing large social networks \cite{Heer_Boyd_2005}, improving recommendation systems \cite{sahebi2011community}, compressing network data \cite{boldi2010compressing}, and viral marketing \cite{journals/abs-physics-0509039} can be applied on the inferred network, and lead to comparable results with the original underlying network.

In this paper, we propose a new Diffusion Aware Network Inference algorithm, called DANI, which is intended to infer social networks from diffusion information while persevering the community structures. We consider the infection time of nodes in different contagions as observations, and apply community detection algorithms on the inferred network. Moreover, by comparing the detected communities of the actual network with the inferred ones, we show that communities are detected successfully.

The main contributions of this paper are:
\begin{itemize}
\item Inferring networks while preserving their properties: Since it is not feasible to obtain the real network structure, different structural based methods such as community detection algorithms may use the inferred networks as an alternative. The previous network inferring works do not consider the density of edges and their degree distributions in the underlying network. Consequently using community detection methods on them deviates from community detection on the original network. However, other structural properties such as degree distribution and clustering coefficient will be different from the ground truth. In this way, the ranking of nodes will be distributed based on their influence role. The proposed method overcomes this problem by preserving the community structure while inferring the networks.
\item Inferring networks with more connected nodes: Network nodes in inference problems are the active nodes participating in at least one observed cascades. Hence, any active node must have at least one connected link providing the required infrastructure for cascades to spread through that link. Compared to the previous works, the proposed method can produce an inferred network with more connected links for each node as existed in the underlying network. In other words, our method produces less isolated nodes compared to the previous methods.
\item More scalability: The proposed method offers lower run time compared to the state of the art methods while maintaining high precision. This property leads our work to have more scalability that is a main limitation of the pervious works.
\item Good performance independent of the network structure: Many of the state of the art previous works such as \cite{Netinf:2010,Netrate11,journals/abs-1205-1671} have good performance on the tree-based networks \cite{Fastinf11}, while the performance of the  proposed method is not limited to any type of structure.
\end{itemize}

The rest of this paper is organized as follows. In Section \ref{sec:preliminaries}, we introduce the preliminaries on the concepts of community and diffusion and explain their interdependencies in order to ascertain our concerns about preserving community structures during the inference process. Related works are studied in Section \ref{sec:relatedwork}. In Section \ref{sec:proposedmethod}, we define notations, formalize our problem and explain the proposed method. Experimental results on synthetic and real datasets are presented in Section \ref{sec:experimental}. Finally, we conclude the paper and discuss future works in Section \ref{sec:conclusion}. 

\section{PRELIMINARIES ON COMMUNITIES AND DIFFUSION}
\label{sec:preliminaries}

A social network is usually represented by a graph $G(V,E)$. Nodes ($V$) and links ($E$) of these graphs stand for people and their interactions, respectively. The role that different nodes and links play in the network processes such as diffusion, varies and can be described as \cite{Chen2007h,journals/abs-1110-5813,fortunato2010}: 
\begin{itemize}
\item \textit{Intra-community links} that connect nodes in the same community. Such nodes interact frequently; therefore, these links are known as strong ties. Nodes that only have links to other nodes in their own community are called core nodes. The core nodes mostly affect their cohorts in a community more than other members of the network.
\item \textit{Inter-community links} that connect nodes of different communities. Due to low interactions between such nodes, they are called weak ties. Nodes, which have at least one inter-community link, are called boundary nodes. Such nodes play an important role in information dissemination over different communities.
\end{itemize}
\begin{figure}[]
\centering
\includegraphics{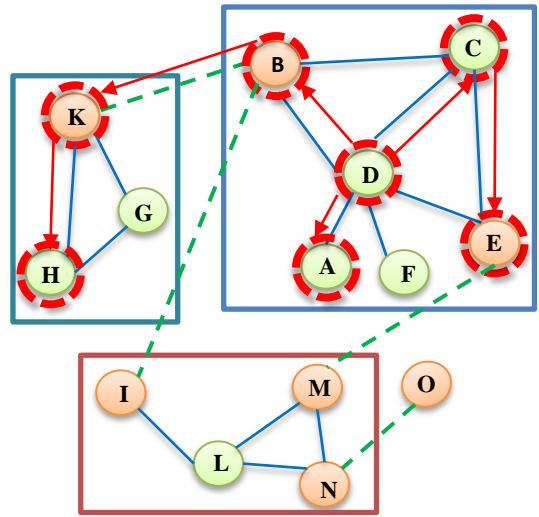}
\begin{center}
\caption{\label{fig:cascade} (Color online) Three communities are separated by rectangles around them. Strong ties (blue solid undirected lines) form a community, and weak ties (green undirected dash lines) act as bridges between different communities. Core nodes (green circles: $\{A,C,D,F,G,H,L\}$), and boundary nodes (orange circles: $\{B,E,I,K,M,N,O\}$) are shown. The directed path (red directed lines and bold nodes) shows a typical cascade over the network.}
\end{center}
\end{figure}

Analyzing the interactions between community members is an important issue in social networking \cite{FuzzySystemsChen}. In fact, the probability that a person would reject or accept a contagion depends on the decision of his neighbors about that contagion. At the same time, this probability affects the future decisions of neighborhood nodes in the same community \cite{scripps2007roles}. By strong intra-community links, members in a community are exposed to a contagion during a small interval from several paths. In contrast, due to weak inter-community links between communities, propagation of a contagion among them takes place with lower probability and pace. 

Another interesting finding is that in many cases the impact of weak ties on a network process (i.e. acceptance rate of a new product) is more powerful than strong ties \cite{abs-1201-4145,goldenberg2001talk}. We try to explain the above facts through an example. A network with three communities is illustrated in Fig.~\ref{fig:cascade}. Suppose that node $D$ gets infected by a contagion. Other nodes in the community of node $D$ will be exposed to this contagion and become infected with high probability due to similar interests and actions among them. Thus, the rate and extent of diffusion can be high within a community. But how does the contagion disseminate over the network? For this purpose, the contagion must spread out of the community by a boundary node such as $B$, and infect another boundary node like $K$ from a neighboring community, and this only happens if there exists at least one inter-community link such as $(B,I)$ or $(B,K)$. Therefore, weak ties have a critical role in the diffusion process over the network. Since interaction on weak ties happens with lower probability, the diffusion process may stop. In other words, diffusion and community may act against each other. Communities may prevent a cascade, and in most cases when a cascade stops, a community can be detected \cite{easley2010,BarbieriBM13}. 

Therefore, it is necessary to understand the relation between the community structure and diffusion behavior of nodes in a network in order to infer the network correctly.   

\section{RELATED WORK}
\label{sec:relatedwork}

As discussed in the previous sections, we would like to infer a network by utilizing diffusion information while taking into account its community structure.  Consequently, the most related research areas to our work are:
\begin{itemize}
\item Diffusion aware network inference: The methods that infer networks by observing the diffusion network.
\item Diffusion aware community detection: The methods that utilize the diffusion process during community detection. 
\end{itemize} 
In this section, we will survey and explain the state of the art methods in each research area.

\subsection{Diffusion based Network Inference}
\label{sec:Diffusion based Network Inference}

This category of research tries to infer the edges of a network by using cascade information, which in most cases is the infection time of nodes in different cascades \cite{Netinf:2010,nipsMyersL10,EslamiRS11,Netrate11,Fastinf11,journals/abs-1205-1671}. \cite{journals/sigmod/GuilleHFZ13} introduces a comprehensive survey on previous works in this area. In the following, we describe some of the most important works in this category.

NETINF models each cascade as a tree on the network and uses an iterative algorithm to infer the network from these trees by optimizing a sub-modular function \cite{Netinf:2010}. Despite its high inference accuracy, it suffers from high runtime which is a serious problem in real datasets. In addition, the performance of NETINF decreases when the structure of the network being studied is not consistent with the tree model \cite{Fastinf11}.

NETRATE is an improvement over NETINF. It assumes that cascades occur at different rates and temporally infers heterogeneous interactions with different transmission rates, which is closer to reality \cite{Netrate11}.

CoNNIE improves NETINF by adding an optimal and robust approach which uses prior probabilistic knowledge about the relation between infection times \cite{nipsMyersL10}.

Similar to NETINF, MultiTree models cascades with the trees, but considers all the possible trees. By this approach, its accuracy is higher than NETINF, NETRATE and CoNNIE when low number of cascades is available. Although, its running time is several orders lower than other tree-based algorithms, but still it is not scalable. \cite{journals/abs-1205-1671}.  

DNE models diffusion as a Markov random walk and tries to define a weight for each link according to the difference between the infection times of connected nodes in different cascades \cite{EslamiRS11}. Its approach is independent of diffusion models. FastINF is another method which uses the same approach to define the weight of links \cite{Fastinf11}. Both of these algorithms have a low runtime and work well on the networks that do not have meaningful community structures.

The overall goal in such studies is to obtain the maximal number of edges of the underlying network. The main drawbacks of this category of studies are as follows.  

Algorithms like NETINF and others based on it, which use a spanning tree approach, maintain the structures, but are too slow. These methods have low performance on networks with non-tree structures.

Others like DNE and FastINF distinguish no preference between different types of the extracted edges. Despite their high accuracy, they do not consider the network community structure. Running community detection algorithms on the output of these methods leads to different communities from underlying network, because they do not recover some weak inter-community or strong intra-community links. Therefore, few large communities or many single members are seen in the networks inferred by them. 

Our goal is to infer the network by utilizing cascade information such that the community structure of underlying network is preserved. In this regard, we desire to maintain network community structure through the inference process by finding more intra-community links to recognize relatively dense communities in the network.

\subsection{Diffusion based Community Detection}
\label{sec:Diffusion based Community Detection}

Many community detection algorithms have been proposed in recent years. The main assumption of these algorithms is the availability of the underlying network topology (nodes and links between them). However, in many cases, the network topology (i.e. friends, followers or following of users) is latent which causes the accuracy of traditional community detection algorithms to reduce. 

Different interpretations of diffusion are used in community detection algorithms. The most prevalent definitions are described in the following.  1) Artificial diffusion:  It is mainly produced by well-known diffusion models \cite{kempe03maximizing} such as Independent Cascade (IC) model \cite{conf/kes/SaitoNK08} and Linear Threshold (LT) model \cite{chen2010scalable}. Moreover, some studies have introduced new behaviors for diffusion \cite{10.1109/CyberC.2010.57}.  2) Real diffusion: It corresponds to a foreign traceable process on the network that appears as part of a dataset information. For example, in news cascade over Twitter (\url{http://twitter.com}), if user $u$ tweets a post, his follower node $v$ will see this action on his page. In general, we say node $v$ is exposed to a contagion, and can retweet. Another example is voting stories in Digg (\url{http://digg.com}) which allows users to post news that can propagate according to the rating of users.

We categorize community detection algorithms according to their diffusion type as follows:

\textit{Detecting communities by artificial diffusion}: 
These studies create artificial diffusion over the network and offer a community detection algorithm that uses this diffusion as part of its procedure. In multi-agent approaches, information such as color, is exchanged between nodes via the edges according to different diffusion models. At the end, nodes with the same colors are detected as a community \cite{10.1109/ASONAM.2009.23}. Some works use label propagation in order to define communities. Nodes adopt the maximum label of their neighbors in an iterative process, and at last nodes with identical labels form communities \cite{Raghavan2007Near,Costa2004c,leun08}.  In another research, the number of common neighbors between nodes is used to define the weight of edges as a similarity measure \cite{FuzzySystemsChen}. By considering this assumption, another work introduces a new comportment named diffusion, which combines weighted edges in a hierarchical method to detect communities \cite{10.1109/CyberC.2010.57}. 

\textit{Detecting communities by real diffusion}: 
The only major work in this category proposes a stochastic generative model named CCN that uses both the complete network topology and diffusion information to extract communities \cite{BarbieriBM13}. This model does not consider any special assumption about link formation, arrival probability of a contagion at each node, cascades models over the network, and rather tries to model them with a random process. Despite good features such as overlapping community detection, this method suffers from high running time complexity. In \cite{DBLP:conf/icdm/BarbieriBM13}, the authors expand CCN by considering that the underlying network is not available. This method adopts a mathematical model similar to CCN, and uses an independent cascade assumption to introduce a model called C-IC, and also uses NETRATE to introduce a model called C-Rate. The output of this method is the network communities without producing the links of network.

In general, low performance in the absence of network topology and being restricted to undirected networks are the main drawbacks of previous community detection algorithms. In contrast to prior algorithms, our method assumes the network topology to be latent, and aims to infer the network utilizing infection time of nodes, which is available in most realistic cases. In addition, our method performs well for both directed and undirected networks.

Network inference is the main goal of this article which is more related to works explained in the \ref{sec:Diffusion based Network Inference} category. Community detection algorithms that are based on diffusion concepts are not relevant to the proposed method in this paper, because they assume that the complete network topology is accessible. On the other hand, \cite{DBLP:conf/icdm/BarbieriBM13} which is the only community detection work that does not consider the aforementioned assumption, utilizes a network inference approach (\cite{Netrate11}). 

\section{PROPOSED METHOD}
\label{sec:proposedmethod}

In this section, we first present the notations used in this paper, and then introduce the proposed algorithm.

\subsection{Notations and Definitions}
\label{sec:Notations and Definitions}
Let $\mathcal{G}=(V,E)$ represent a network where $V$ contains the set of nodes, and $E$ contains the directed or undirected links between nodes. We assume $k=|E|$ (number of links) is known, while no more information about the links is available. We assume to have a set of different contagions as different cascades,  $C=\{c_1,c_2,\ldots,c_M\}$. For each cascade $c_i$, we observe the infection time of nodes and have pairs of node ID and infection time, $c_i=\{(v_0,t_0),(v_1,t_1),\ldots,(v_n,t_n)\}$. During the spread of a contagion some nodes may not get infected. An infinity infection time is assigned to such nodes.

The propagation path of a contagion is hidden. In other words, we do not know by which node, a node gets exposed to a contagion. It is assumed that a node can get infected with a certain contagion just by one other node. 



In general, infection times form a continuous sample space which makes their interpretation somewhat difficult. In addition, the range of infection time may vary for each contagion. For example, the order of infection time maybe in seconds for cascade $C_a$, but in order of hours for another cascade $C_b$. Therefore, if we use the real value of the infection times, we will not infer the links that $C_b$ spreads over the network; since the time differences in $C_b$ are much larger than $C_a$, the information obtained from $C_b$ would be ignored. To overcome this problem, we define a function that maps these continuous values to a set of discrete values and refer to the corresponding random variable as \textit{infection label}. Here, our goal is to discover the links of a static network structure from the observed cascades.
We do not consider the speed of cascades and expect two different cascades to spread in the same manner for a same path on the graph. We also ignore the impact of cascades on each others. Therefore, instead of time of propagation, the sequence of propagation is important and we achieve this representation by mapping time values to infection sequence order per each cascade. The proposed infection time transformation can be described as follows.

\begin{enumerate}
\item For each cascade $c_m$, sort the elements according to their infection time ($t_u$) in an ascending order. The resulting sorted set is named $S_m$. Elements with infinity infection time are omitted from $S_m$.
\item For each node $u$, assign a label equal to its position in $S_m$, and replace its infection time with that label.
\begin{equation}\label{eq:assignlabel}
L({t_u}) = \left\{ {\begin{array}{*{20}{c}}
{index({S_m}({t_u}))}&{if({t_u} \prec \infty )}\\
0&{else}
\end{array}} \right.
\end{equation}
\end{enumerate}

This way, each cascade $c_m$ is a vector containing pairs of nodes and infection labels that are sorted in an ascending order according to their infection labels, as depicted in Eq.~(\ref{eq:cascadevector}). From now on, we call this vector the \textit{cascade vector}, $CV_m$:
\begin{equation}\label{eq:cascadevector}
CV_m=\{(v_0,L(t_0)),(v_1,L(t_1)),\ldots,(v_n,L(t_n))\}
\end{equation}
For simplicity we may use $L(t_i )$ and $L(v_i )$, interchangeably.

\subsection{Problem Formulation}
\label{sec:Problem Formulation}

We assume, the underlying network $\mathcal{G}= (V,E)$ is unknown, and the only observation is the set of cascades ($C$). If each node of network participates in at least one cascade, the set of nodes $V$ can be extracted from the observation. By considering all the links between any pairs of these nodes $\{(u,v)|{u,v} \in {V}\}$, we have a complete graph ${G}$. Our goal is to obtain a weighted adjacency matrix $A = \{{\alpha_{uv}|u,v} \in {V},{u} \ne {v}\}$ for graph $G$ such that each element ${\alpha _{uv}}\ge0$ of $A$ is calculated considering the community structure preservation property. We utilize the same assumptions and maximum likelihood estimation as in \cite{Netinf:2010,Netrate11,Fastinf11,journals/abs-1205-1671}. These assumptions are used in the following to simplify the maximization process. 
We would like to assign weight for each edge of complete graph ${G}$ that maximizes the following likelihood function.

\begin{equation}\label{eq:MLEG2}
{G^{'}} = \arg {\max _{G}}P(C|G)
\end{equation}

Then, we extract a sub-graph $G^*$ of $G^{'}$ that has the $k$ most weighted edges as probable edges of the latent network ($k$ is the number of underlying network links which is given as input).

%
%
%
%
For simplicity, we assume the spread probability of cascades to be independent and identically distributed (i.i.d) over the graph. Therefore, the probability of observing the cascade set over the networks is:
\begin{equation}\label{eq:CascadeOverNet}
P(C|G) = \prod\limits_{i = 1}^{i = m} {P({c_i}|G)} 
\end{equation}

%

Graph $G$ is composed of set of edges $(u,v)$, and we assume that all edges are mutually independent:

\begin{equation}\label{eq:GIndependence}
P(G) =\prod\limits_{(u,v) \in G} {P(u,v)}
\end{equation}

%

Therefore, the probability that cascade $c_i$ spreads over graph $G$ is given by Eq.~(\ref{eq:CiconditionalGTotal}), assuming that the joint probability of edges are conditionally independent:
\begin{eqnarray}\label{eq:CiconditionalGTotal}
P({c_i}|G) = P({c_i}|{\{ (u,v)\} _{(u,v) \in E}}) = \frac{{P({{\{ (u,v)\} }_{(u,v) \in E}}|{c_i}) \times P({c_i})}}{{P({{\{ (u,v)\} }_{(u,v) \in E}})}}\nonumber\\
 = \frac{{\prod\limits_{(u,v) \in E} {P((u,v)|{c_i})}  \times P({c_i})}}{{\prod\limits_{(u,v) \in E} {P(u,v)} }}\nonumber\\
 = \frac{{\prod\limits_{(u,v) \in E} {\frac{{P({c_i}|(u,v))P(u,v)}}{{P({c_i})}}}  \times P({c_i})}}{{\prod\limits_{(u,v) \in E} {P(u,v)} }}\nonumber\\
\to P({c_i}|G) = P{({c_i})^{ - (|E| - 1)}} \times \prod\limits_{(u,v) \in E} {P({c_i}|(u,v)){\rm{ }}} 
\end{eqnarray}
Where $P({c_i})$ is the prior probability for cascade formation, and we assume it is constant for all cascades. Therefore, we can approximate the probability of cascade $c_i$ propagating over $G$ by Eq.~(\ref{eq:CiconditionalG}):

\begin{eqnarray}\label{eq:CiconditionalG}
P({c_i}|G) \propto \prod\limits_{(u,v) \in G} {P({c_i}|(u,v))} 
\end{eqnarray}
hence, by using Eq.~(\ref{eq:CiconditionalG}), we can rewrite Eq.~(\ref{eq:CascadeOverNet}) as:
\begin{eqnarray}\label{eq:CconditionalG}
P(C|G)\propto\prod \limits_{i = 1}^{i = m}{\prod \limits_{(u,v) \in G}{P(c_i|(u,v))}}\nonumber\\
\propto \prod\limits_{(u,v) \in G}{\prod\limits_{i = 1}^{i = m}{P({c_i}|(u,v))}}
\end{eqnarray}
Applying log function on both sides of Eq.~(\ref{eq:CconditionalG}), yields:
\begin{equation}\label{eq:LogCconditionalG}
\log (P(C|G)) \propto \sum\limits_{(u,v) \in G} {\sum\limits_{{c_i} \in C} {\log (P({c_i}|(u,v))} } 
\end{equation}
where $P({c_i}|(u,v))$ represents the probability that cascade $c_i$ spreads through edge $(u,v)$. We try to define the probability $P({c_i}|(u,v))$ in terms of observed information from the hidden network structure. 
We interpret $log(P(c_i|(u,v)))$ as the weight of edge $(u,v)$, and call it ${\alpha _{c_i}}$. We try to estimate this weight for each presumable link in the network. In the rest of paper, we introduce the DANI algorithm in which ${\alpha _{c_i}}$ is computed according to the relations between community structure and diffusion information.

\subsection{The DANI Algorithm}
\label{sec:DANI Algorithm}

As mentioned before, our goal is to define a weight $P(u,v)$ for each edge $(u,v)$ that represents its existence probability in the underlying network. For this purpose, DANI first utilizes diffusion information and estimates the weight of all potential edges. At the second step, DANI considers the relation between community structures and diffusion behavior and assigns another weight according to the joint behavior of nodes. Finally, both of these weights are taken into account simultaneously, and the existence probability of each edge is computed. In summary, we follow three steps: using information diffusion network, maintaining the community structure, and the combination of these steps.

\textbf{Diffusion Information:} From each cascade vector $CV_m$ we can construct a probable directed graph. Each node $u$ in this graph can get infected by a node $v$, if $L(v) \prec L(u)$. The probability of existence of edge $(u,v)$ in this graph depends on the difference between the infection label of $u$ and $v$, such that more difference between the infection labels of these nodes in $CV_m$, decreases the participation probability of link $(u,v)$ in the contagion \cite{Netinf:2010}:
\begin{equation}\label{eq:LinkParticipationProbability}
P(u,v) \propto \frac{1}{{\Delta (L(u),L(v))}}
\end{equation}

\begin{figure}[]
\begin{center}
\includegraphics[width=0.99\textwidth]{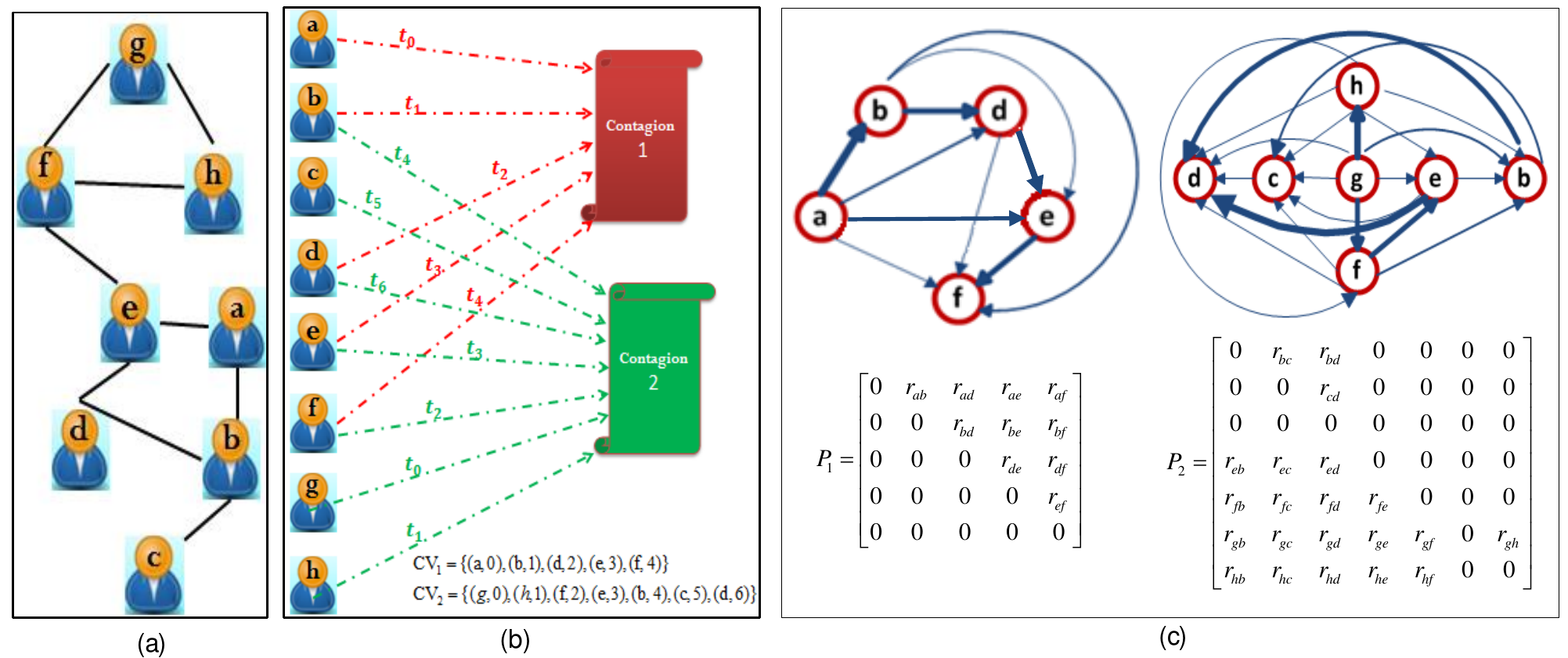}
\caption{\label{fig:MarkovChainModel} (Color online) (a) A network topology. (b) Cascades over network: a bipartite graph with network members in one side and contagions on the other. Infection of each user with a contagion is represented by an edge from it to that contagion, where the infection time is the edge label. The cascade vector calculated by the proposed method is also considered. (c) Markov chain and matrix for each cascade: probable directed graph with the DANI algorithm. Edge probabilities are marked with thick and thin lines. The Markov transition matrix per cascade is obtained from these graphs as shown here.}
\end{center}
\end{figure}

Fig.~\ref{fig:MarkovChainModel}(a) illustrates a typical network and two of its distinct contagions. Their cascade vectors and corresponding probable graphs (simply referred to as graph) are shown in Fig.~\ref{fig:MarkovChainModel}(b) and Fig.~\ref{fig:MarkovChainModel}(c), respectively.

Each cascade vector $CV_m$ can be modeled as a discrete time stochastic process where nodes stand for independent states and the sequence of nodes in the cascade vector $CV_m$ represent the state of the process at discrete times. The transition probability between different states (nodes), or in other words, the contagion transmission probability from a node to another node is equivalent to the weighted link between them. 

At the present time the contagion transmission probability of each node (state) depends on probability of being at its neighboring node in the immediate preceding period, and is independent of its pervious time states or the traversed nodes. This property is called the \textit{Markov Property}. Hence, it can be thought that the states of former times are all accounted by incorporating the state of the last immediate time \textit{t}. Assuming the First Order Markov Property to hold, we have:
\begin{equation}
\begin{array}{c}
Pr[{X_{t + 1} ={x_{t + 1}}{\rm{|}}{X_t} = {x_t},{X_{t - 1}} = {x_{t - 1}}, \ldots ,{X_0} = {x_0}]}\\
=Pr[{{X_{t + 1}} = {x_{t + 1}}{\rm{|}}{X_t} = {x_t}]}
\end{array}
\end{equation}


This stochastic process that exhibits the Markov property is called a Markov Chain (MC). The number of states in our Markov chain is equal to the number of nodes, and each cascade vector $C{V_m}$ is a finite Markov chain \cite{LevinPeresWilmer2006}.

In order to define the Markov chain, we must compute its transition matrix. Each element of this matrix represents the existence probability of an edge in the network. As shown in Fig.~\ref{fig:MarkovChainModel}(c), our goal is to find the transition matrix for each chain by utilizing the corresponding cascade information. 

At this point, we need to find the relation between the difference of the infection labels of two nodes, and the probability that an edge exists between them. As discussed before, this probability has an inverse relation with the difference of their infection labels. Two approximations are used in the previous works:
\begin{enumerate}
\item The probability of an edge only depends on the difference of the infection label of nodes connected by it. In other words, for all pairs of nodes with the same values of infection label difference, the existence probability is the same \cite{Netinf:2010}. 
\item Consider a cascade vector such that there exists a set of nodes ($D$), before $v$ in this vector. The probability that $u$ infected $v$ when $D =\{u\}$, is higher than the situation where $|D|\succ2$, because a node $z \in D$ with ${t_z} \prec {t_v}$ may have infected $v$. This will reduce the probability that $v$ has been infected by $u$. Moreover, the authors in \cite{EslamiRS11} proved that in addition to the difference of infection times, the number of infected nodes before node $v$, namely $|D|$, also affects the edge weights.
\end{enumerate}

Considering the above points, the existence probability of an edge between nodes $u$ and $v$ in the cascade vector ($L(v) \succ L(u)$), using the infection label is:
\begin{equation}\label{eq:Diffusionbasedprob}
d_{u,v} = \frac{1}{{L(v)\times(L{(v) - L(u)})}}
\end{equation}

We use $d_{u,v}$ to construct the transition matrix. The transition matrix $P$, is a stochastic random matrix with the following features:
\begin{enumerate}
\item Each element in $P$ satisfies the following condition:
\begin{equation}
0 \le {P_{ij}}\enspace  \enspace i,j=1,2,...,n
\end{equation}

\item The sum of each row is equal to one:
\begin{equation}
\forall i:\sum\limits_{j = 1}^{j = n} {{P_{ij}}}  = 1
\end{equation}
\end{enumerate}

To satisfy these conditions, the two-dimensional probability matrix should be normalized. Dividing each element of this matrix by the sum of its corresponding row, the Markov chain transition matrix ($P_{c_i }$), for cascade ($c_i$) is obtained:
\begin{eqnarray}\label{eq:CascadeTransitionMatrix}
{P_{{c_i}}}\left( {u,v} \right) = \frac{{{d_{u,v}}}}{{{d_u}}}\nonumber\\
{d_u} = \mathop \sum \limits_{y \in V} {d_{u,y}}
\end{eqnarray}

\textbf{Community Structure Information:} As previously mentioned, community members have similar interests and behaviors, and diffusion is more likely to spread among them \cite{easley2010,Eftekhar:2013,BarbieriBM13,DBLP:conf/icdm/BarbieriBM13}. Intuitively, they appear in same cascades more than nodes which are in different communities. Thus the number of cascades passed between node $u$ and $v$ can be a sign of coherence between these nodes. If $u$ and $v$ are community members, they become infected jointly in more cascades compared to the cascades that only one of them has been infected. We use the participations of pairs of nodes in different cascade as a mark for finding the similarity in their behaviours.

On the other hand, consider two boundary nodes that connect adjacent communities. These nodes may appear in same cascades due to the bridge role of weak ties in the diffusion process.  However, the frequency of such cases is negligible compared to the number of cascades that every single one of them has participated in. 

\begin{figure}[]
\begin{center}
\includegraphics[width=0.4\textwidth]{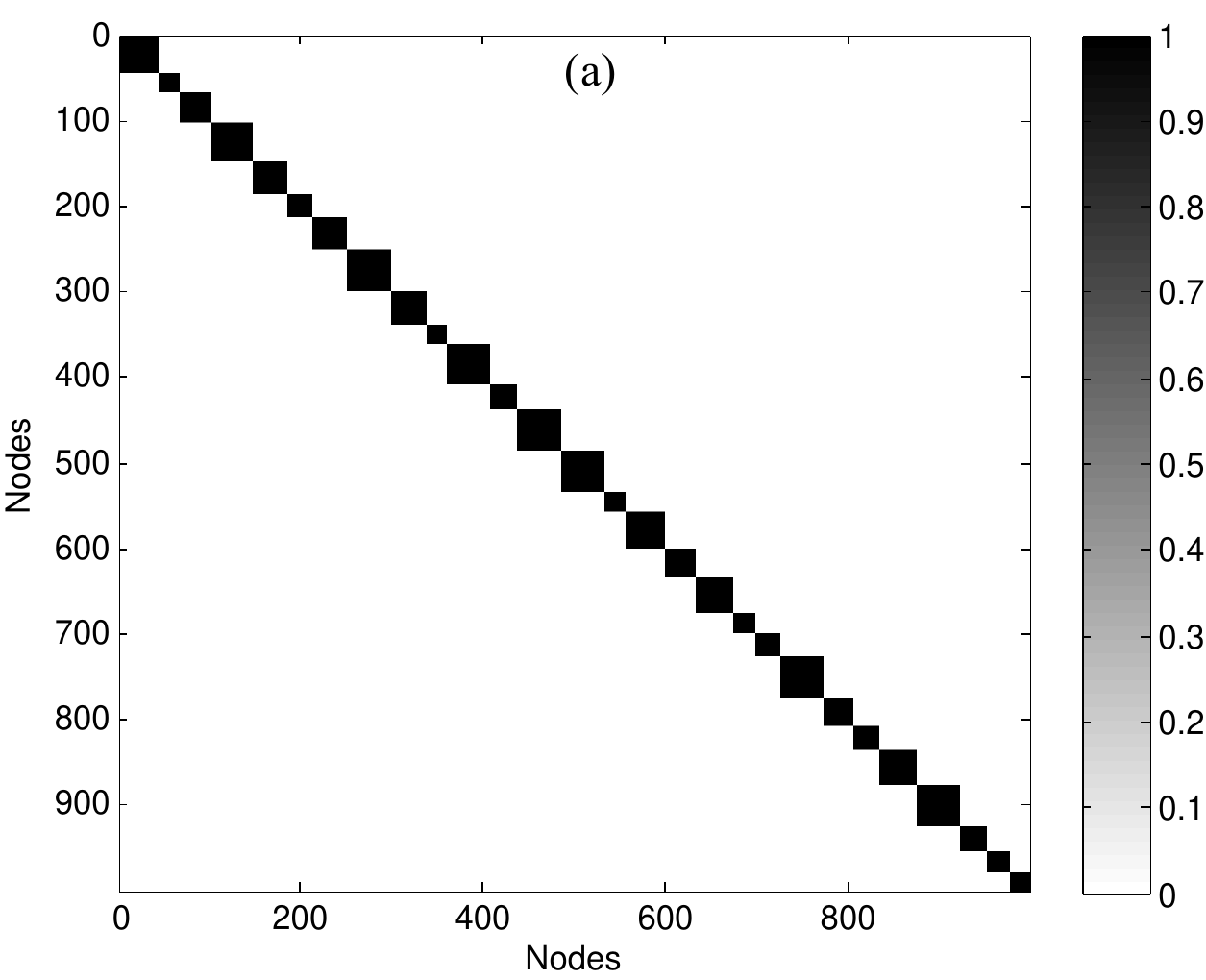}
\includegraphics[width=0.4\textwidth]{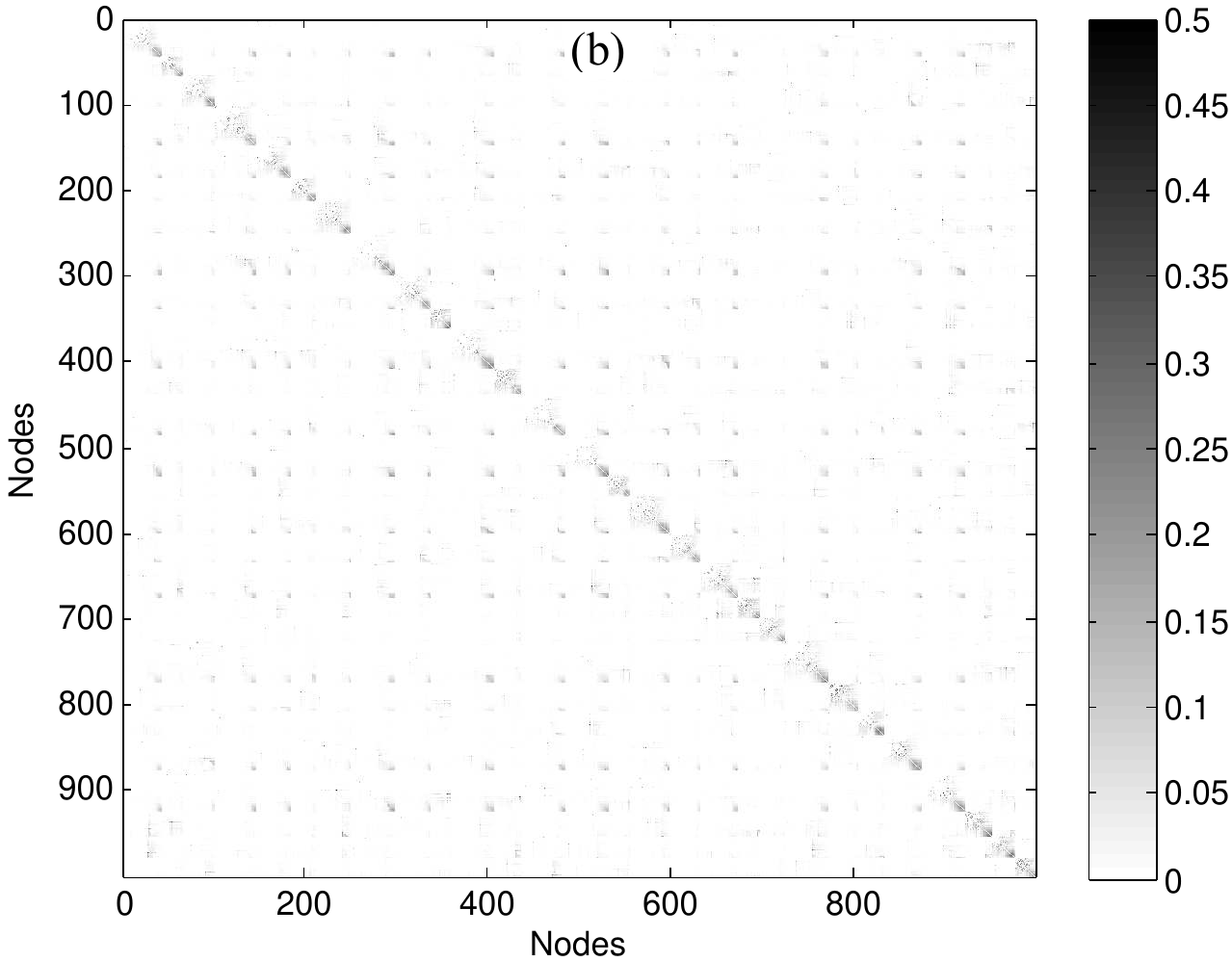}
\caption{\label{fig:CommunitySimilarity} The color map between any pair of nodes. (a) Community structure: If two nodes are in a community, the pixel is black, otherwise it is white. Each block represents a specific community. (b) Similarity: We measured similarity between any pair of nodes by Eq.~(\ref{eq:edgeweight}). Dark pixels indicate more similarity. The pixel of similarity matrix is darker wherever the community matrix is black, leading to the conclusion that communities can be specified by using the node-node diffusion based similarity.}
\end{center}
\end{figure}

According to the above justifications, and the goal of inferring the links for each pair of node $(u,v)$, we define a function $\psi \left( {u,v} \right)$ that assigns a weight to edge $(u,v)$ by considering the nodes behaviourals based on the cascades information. In other words, this weight is a node-node similarity based on the community structure that had impacted the diffusion information. The formula for $\psi \left( {u,v} \right)$ is similar to the Jaccard index \cite{jaccard}:
\begin{equation}\label{eq:edgeweight}
\psi \left( {u,v} \right) = \frac{{\left| {{In(u)\bigcap {In(v)} } } \right|}}{{\left| {{In(u)\bigcup {In(v)} }} \right|}}
\end{equation}
where $In(u)$ is the set of cascades that $u$ has participated in:
\begin{equation}
In(u) =
\left \{ {\begin{array}{*{20}{c}}
{\bigcup\limits_{i =1}^M \{i\}} &{if(u \in C{V_i})}\\
{ {\emptyset}} & {else}
\end{array} } \right.
\end{equation}
For computing the numerator of Eq.~(\ref{eq:edgeweight}), $u$ should appear earlier than $v$ when both participate in each $CV_i$.
The above node-node similarity can be used to detect the communities.  The main idea of many structured-based community detection algorithms is grouping nodes based on the fact that when the similarity of two nodes is large, they have more chance to be in a community \cite{du2007community,weng2013virality}.  We also provided a data driven study to show the relation between Eq.~(\ref{eq:edgeweight}) and the aforementioned concept of community. To this end, we utilized an LFR network with 1000 nodes, 7692 links, 28 communities and 20000 cascades (which is described and analyzed in Section \ref{sec:experimental}). We changed the name of nodes such that nodes of a community are consecutively listed. For each pair of nodes, the color map community matrix and similarity matrix in Fig.~\ref{fig:CommunitySimilarity} show that nodes with more similarity are in the same community.

Since overlapping nodes are highly influential nodes \cite{6724361}, this formula also works for overlapping communities. Because by using this equation, the overlap nodes will take part in more common cascades and consequently gain large values. 

\textbf{Total Information:} The probability of direct contagion propagation between two nodes would increase, if the difference of their infection times decreases. In other words, the existence probability of edge $(u,v)$ has a direct relationship with $P_{{c_i}}(u,v)$. From other perspective, larger values of $\psi(u,v)$ correspond to more probability that $u$ and $v$ are cohorts, and hence indicating edge $(u,v)$ to be an intra-community link. Therefore, direct impact of $\psi(u,v)$ on the existence probability of edge $(u,v)$ will lead to infer the intra-community links that are the important links in the community detection step. 
We define the probability weight of an edge at each cascade $c_i$ as:
\begin{equation}\label{eq:APLPHA}
\alpha _{c_i}({u,v}) = P_{c_i}(u,v) \times \psi (u,v) 
\end{equation}
where $P_{c_i } (u,v)$, is the element $(u,v)$ in the Markov transition matrix of the cascade $c_i$. \\\\

To mathematically justify the above equation (Eq.~(\ref{eq:APLPHA})), let us consider the following definition. According to \cite{Malmgren:2008}, independent cascades over the network are Poisson point processes that follow the exponential distribution \cite{Netinf:2010}:
\begin{equation}\label{eq:CascadeProbability}
P({c_i}|(u,v)) \propto e^{\frac{{ - ({t_v} - {t_u})}}{\theta_{u,v} }}
\end{equation}

Then, the probability of cascade ${c_i}$ spreading over an edge $(u,v)$ has an inverse relation with the difference between infection times (infection labels) of node $u$ and node $v$, and the parameter $\theta_{u,v}$. According to discussion in the "Diffusion Information" section the authors in \cite{EslamiRS11} proved that a cascade spread over edge $(u,v)$ does not only rely on the infection label difference of nodes $u$ and $v$, but also the set of nodes $(D)$ that has been infected before activation of node $v$. Therefore, $P({c_i}|(u,v))$ is given by:
\begin{eqnarray}\label{eq:FunctionCascadeProbability}
P({c_i}|(u,v)) \propto exp({\frac{-1}{{({L(t_v)}({L(t_v)} - {L(t_u)}))^{-1}}\times \theta_{u,v} }})
\end{eqnarray}

By substituting Eq.~(\ref{eq:FunctionCascadeProbability}) in Eq.~(\ref{eq:LogCconditionalG}) we obtain:
\begin{eqnarray}\label{eq:TotalLogMLE1}
\log \left( {P\left( {C{\rm{|}}G} \right)} \right) \propto \mathop \sum \limits_{\left( {u,v} \right) \in G} \mathop \sum \limits_{{c_i} \in C} log(exp(\frac{-1}{{({L(t_v)}({L(t_v)} - {L(t_u)}))^{-1}}\times \theta_{u,v}}))
\end{eqnarray}

Considering the logarithm of maximum likelihood function $\arg {\max}(P(C|G))$ for Eq.~(\ref{eq:TotalLogMLE1}) we obtain:

\begin{eqnarray}\label{eq:MinMaxLogMLE}
{G^*}\propto \arg {\max} ( \mathop \sum \limits_{\left( {u,v} \right) \in G} \mathop \sum \limits_{{c_i} \in C} (\frac{-1}{[{L(t_v)}({L(t_v)} - {L(t_u)})]^{-1}\times \theta_{u,v}}))\nonumber\\
\propto \arg {\min} ( \mathop \sum \limits_{\left( {u,v} \right) \in G} \mathop \sum \limits_{{c_i} \in C} (\frac{1}{[{L(t_v)}({L(t_v)} - {L(t_u)})]^{-1}\times \theta_{u,v}}))\nonumber\\
\propto \arg {\max} ( \mathop \sum \limits_{\left( {u,v} \right) \in G} \mathop \sum \limits_{{c_i} \in C} ({[{L(t_v)}({L(t_v)} - {L(t_u)})]^{-1}\times \theta_{u,v}))
}\end{eqnarray}

In Eq.~(\ref{eq:MinMaxLogMLE}), the first term is identical to relation we extracted from diffusion information, and the second term corresponds to the information about link that can map to the relation extracted from community structure information:

\begin{eqnarray}\label{eq:MappingArgMax}
{G^*}\propto \arg {\max} ( \mathop \sum \limits_{\left( {u,v} \right) \in G} \mathop \sum \limits_{{c_i} \in C} \underbrace {({[{L(t_v)}({L(t_v)} - {L(t_u)})]^{-1}}}_{P_{c_i(u,v)}}\times {\underbrace {\theta_{u,v})}_{\psi ({u,v})}})
\end{eqnarray} 

Then, according to our model the Maximum Likelihood Estimator (MLE) is given by:
\begin{eqnarray}\label{eq:TotalLogMLE}
{G^*}\propto \arg {\max} (\mathop \sum \limits_{\left( {u,v} \right) \in G} [\mathop \sum \limits_{{c_i} \in C} {P_{{c_i}}}\left( {u,v} \right)] \times \psi \left( {u,v} \right))\nonumber\\
\propto \arg {\max} (\mathop \sum \limits_{\left( {u,v} \right) \in G} {P_C}\left( {u,v} \right) \times \psi \left( {u,v} \right))
\end{eqnarray}
where $P_C $ is the sum of the transition matrix of all cascades, preserving the Markov matrix properties, and it stands for the total transition matrix of the Markov chain. 

The DANI pseudo-code is presented in Algorithm \ref{alg:DANI}, where $P_{temp}$ is used to keep the sum of all transition matrices of cascade vectors.  Then, $P_{temp}$ is normalized to obtain the stochastic transition matrix $P_{C}$, based on the diffusion information. In the next step, by considering the community structures, $\psi (u,v)$ is calculated and used along with $P_{C}$ to produce the inferred network adjacency matrix, $A$. The normalization in the process of computing $P_{C}$ keeps its stochastic transition property and its value within the limits of $\psi (u,v)$ values, and hence their multiplication stays in a fix range for all the possible edges. Finally, we choose $k$ edges of the inferred network with maximum weight to construct $G^*$. 
\begin{algorithm}[]
\caption{DANI Pseudo-Code}
\label{alg:DANI}
\SetAlgoNoLine
\KwIn{Set of cascades over network ($C = \left\{ {{c_1},{c_2}, \ldots {c_M}} \right\}$), Number of network links to be inferred $(K)$}
\KwOut{Inferred network ($G^*$)}
\For{each ${c_m} = \left\{ {\left( {{v_i},{t_i}} \right)} \right\} \in C$}
{
${S_m}$ $ \leftarrow $ sort ${c_m}$ by ${t_i}$\;
$L\left( {{t_i}} \right) = index\left( {{S_m}\left( {{t_i}} \right)} \right)$\;
$C{V_m} = \left\{ {\left( {{v_i},L\left( {{t_i}} \right)} \right)} \right\}$\;
\For{each $u \in C{V_m}$, $v \in C{V_m}$}
{
\If{($L({t_u}) < L({t_v})$)}
{
\textbf{Compute} ${d_{u,v}}$ according to Eq.~(\ref{eq:Diffusionbasedprob})\;}
}
\For{each $\left( {u,v} \right) \in {P_{{c_i}}}$}{
${P_{{c_i}}}\left( {u,v} \right) \leftarrow \frac{{{d_{u.v}}}}{{\mathop \sum \nolimits_{y \in V} {d_{u.y}}}}$\;
${P_{temp}}\left( {u,v} \right) \leftarrow {P_{temp}}\left( {u,v} \right) + {P_{{c_i}}}\left( {u,v} \right)$\;
}
}
\For{each $\left( {u,v} \right) \in {P_{temp}}$}
{${P_{temp}}(u,v) \leftarrow \frac{{{P_{temp}}(u,v)}}{{\sum\limits_{y \in V} {{P_{temp}}(u,y)} }}$ \;
${P_c}(u,v) \leftarrow {P_c}(u,v) + {P_{temp}}(u,v)$\;
}
\For{each $(u, v) \in {P_C}$}{
\textbf{Compute} $\psi (u,v)$ according to Eq.~(\ref{eq:edgeweight})\;
$A(u,v) = {P_c}(u,v) \times \psi (u,v)$\;}

${G^t }$ $\leftarrow$ sort ${A}$ values\;
\For{$i=1$ to $K$}
{{$G^*$} $\leftarrow$ ${G^* }$ $\bigcup {\{ {g_i} \in {G^t}\} }$ \;}
\textbf{Return} {$G^*$}\;
\end{algorithm}

\textbf{Time complexity analysis:} Consider a cascade $c_m$ with $n_{c_m}$ infected nodes. According to Algorithm \ref{alg:DANI}, any sorting has the time complexity of $O(n_{c_m}log(n_{c_m}))$. There are $\left( {\begin{array}{*{20}{c}} {n_{c_m}}\\2 \end{array}} \right)$ pair of nodes $(u,v)$ for calculating $d_{u,v}$ and other normalizations with the time complexity of $O({n_{c_m}}^2)$. Therefore, if we have $M$ different cascades with average number of nodes $\overline {{n_c}}$ as the input, the DANI algorithm has a complexity of $O(M \times {{\overline {{n_c}}}^2)}$. It is worth to mention that the baselines works for the competing methods in our experimental analysis did not provide time complexity analysis for their algorithms, because their methods are heavily dependent to the network structure. Instead they just relied on experimental results for running times, which we also compare in the next section.
MultiTree and Netinf use greedy hill climbing to maximize the corresponding sub-modular function through an iterative approach where an edge is inferred at each step if by adding the edge a marginal improvement is achieved. After finding ${K}$ edges, their algorithms are terminated. In contrast, DANI is free of any iteration and thus outperforms MultiTree and Netinf in running time. 

\section{EXPERIMENTAL EVALUATION}
\label{sec:experimental}

In this section, first we introduce the evaluation criteria for two concepts of network inference and community structures detection. In order to evaluate the proposed method, we apply DANI to synthetic and real datasets and compare our results with other static works under the same conditions. 
Since the main goal of this paper is network inference while preserving the community structure, the community detection methods should work better when used with the proposed method. Therefore, we have selected the state of the art inferring works for our comparison study instead of using different community detection methods which is not necessary and feasible in the context of this paper. The three selected methods for comparison include NETINF (as the first proposed method in this area), DNE (as the fastest method), and MultiTree (as the newest approach with superior performance over the previous tree-based methods based on accuracy and running time).

\subsection{Evaluation metrics}
\label{sec:Evaluation metrics}

As mentioned before, previous methods focus on inferring network links with high accuracy and overlook network properties such as community structure. Moreover, most of these methods suffer from high running time. Considering the best practices, we utilize two types of analysis as follows.

\subsubsection{Network inference:}

The main goal of this area is to obtain the highest similarity between inferred and underlying networks. The most common and important issues in this regard are the number of discovered links, and inference process duration. To this end, the main criteria are as follows:

\paragraph{F-measure:}
Precision and recall are two frequently used metrics, and are defined as:
\begin{eqnarray}
Precision=\frac{{|{E_G}\bigcap {{E_{G^*}}|} }}{{|{E_{G^*}}|}}\nonumber\\
Recall=\frac{{|{E_G}\bigcap {{E_{G^*}}|} }}{{|{E_G}|}}
\end{eqnarray}
where $E_G$ and $E_{G^*}$ represent the set of links in the original and inferred network, and $|E_G|$ is the cardinality of $E_G$. There is an inverse relation between recall and precision. To consider both, their geometric mean, named F-measure (or F-score) is used \cite{makhoul_et_al_1999}.
\begin{equation}
F-measure = \frac{2 \times Precision \times Recall}{Precision+Recall}
\end{equation}

\paragraph{Running time:}
Most of the previous methods require more time to gain high accuracy. Therefore, running time is an important parameter to be considered in evaluation of these methods.

\paragraph{Number of nodes:}
The previous works only focus on the number of correct inferred links. Despite their high accuracy, no connected links maybe detected for some nodes. Hence, these nodes are omitted in the inferred network and consequently would not be assigned to any community. As a result, only a fraction of original nodes, that are presented in the inferred network, is used in evaluations.

\paragraph{Other network properties:} Preserving the clustering coefficient and degree distribution of nodes is another important concern. For each node, we calculate the relative error between the exact value of each measure in the ground truth and inferred network.

\subsubsection{Preserving community structure:}

As mentioned before, our goal is to propose a method to infer networks by preserving community structure of underlying network. In order to compare the methods in this regard, we extract communities of underlying and inferred networks by a well-known community detection algorithm and compare the communities of these two networks in two ways:
\begin{itemize}
\item Comparing members of communities (NMI, PWF). These metrics evaluate the similarity of each pairs of communities in real and inferred networks.
\item Comparing properties of communities (density, conductance and number of communities). The average of these criteria over all communities of a network is utilized to evaluate the differences of community structure for real and inferred networks. Hence, for each criterion $Y$, we  compute $|\bigtriangleup {(Avg(Y_{r}),Avg(Y_{i}))}|$, in which $ Avg(Y_{r})$ and $ Avg(Y_{i})$ are average of criterion $Y$ for all communities in real and inferred networks, respectively.
\end{itemize}

\paragraph{Normalized Mutual Information (NMI):}
NMI is a probabilistic information theoretic metric that compares the similarity of two sets of communities \cite{manning2008introduction}, and is defined as: 

\begin{equation}
NMI(A,B) = \frac{{ - 2\mathop \sum \nolimits_{i = 1}^{{C_A}} \mathop \sum \nolimits_{j = 1}^{{C_B}} {N_{ij}}{\rm{log}}\left( {\frac{{{N_{ij}}N}}{{{N_{i.}}{N_{.j}}}}} \right)}}{{\mathop \sum \nolimits_{i = 1}^{{C_A}} {N_{i.}}log\left( {\frac{{{N_{i.}}}}{N}} \right) + \mathop \sum \nolimits_{j = 1}^{{C_B}} {N_{.j}}log\left( {\frac{{{N_{.j}}}}{N}} \right)}}
\end{equation}
where $N$ is a ${C_A} \times {C_B}$ matrix, and ${C_A}$ and ${C_B}$ are number of communities for original and inferred networks, respectively. Each element of $N$ corresponds to the number of common memberships for each pair for communities. ${N_{i.}}$ (${N_{.j}}$) is the row $i$ (column $j$) summation, and $N$ is the total number of nodes or the sum of all matrix elements \cite{duch-2005-72}. The range of NMI values is between $0$ and $1$. Actually, NMI for the same communities reaches $1$, and $0$ represents the overall difference. However, for the overlapping communities this metric has been modified \cite{journals/abs-1110-2515,Lancichinetti2009}.

\paragraph{Pairwise F-measure (PWF):}
Let $H_G$ and $H_{G^*}$ represent the sets of node pairs in the same community in the underlying network and the inferred network, respectively. PWF measures how many pairs of nodes in a community at original network are cohort in the communities of the inferred network \cite{yang2009bayesian,conf/icde/QiAH12}.

\begin{eqnarray}
Precision = \frac{{|{H_G} \cap {H_{G^*}}|}}{{|{H_{G^*}}|}},{\mathop{\rm Recall}\nolimits}  = \frac{{|{H_G} \cap {H_{G^*}}|}}{{|{H_G}|}}\nonumber
\end{eqnarray}
\begin{eqnarray}
PWF = \frac{{2 \times Precision \times {\mathop{\rm Recall}}}}{{Precision + {\mathop{\rm Recall}} }}
\end{eqnarray}

We considered PWF, because it is not enough to rely solely on NMI \cite{journals/abs-1110-5813}. Moreover, recent studies have discovered modularity \cite{Newman:2006:16723398} is not a reliable measurement \cite{duch-2005-72}. In addition, because of sparsity in real networks, the obtained modularity values are often incorrect. 

\paragraph{Density:}
Density indicates the ratio of actual intra-community links to all possible links in a community \cite{vasudevancommunity}: 
\begin{eqnarray}\label{eq:Denstiy}
\delta (S)=\frac{{\left| {{E_S}} \right|}}{{\left( {\frac{{\left| {{V_S}} \right|\left( {\left| {{V_S}} \right| - 1} \right)}}{2}} \right)}}
\end{eqnarray}
where $V_s$ and $E_s$ represent the set of nodes and links for the community $S$, respectively.

\paragraph{Conductance:}
This criterion presents the ratio between number of inter-community links ($c_S$), and intra-community links ($m_S$) as follows: \cite{Lescovec08communityStructure}.
\begin{equation}\label{eq:Conductance}
\emptyset  = \frac{{{c_S}}}{{\left( {2{m_S} + {c_S}} \right)}}
\end{equation}

Density and conductance are complementary measures and are often considered together. For instance, the density of a k-clique subgraph with many links to the rest of graph is high, while its conductance low. Therefore, these two criteria are often used together to evaluate the candid methods. In this paper, we desire to have more similarities in density and conductance among original and inferred networks.

\paragraph{Number of communities:}
A measure called ``Accuracy in the number of communities'' was introduced in \cite{conf/icdm/YangL12a} to determine whether the inferred network has the same number of communities as the original network:

\begin{equation}
NC = \frac{{||{CO_G}| - |{CO_{G^*}}||}}{{|{CO_G}|}}
\end{equation}
where $|CO_G|$ and $|CO_{G^*}|$ are the cardinality of the set of communities for original and inferred networks, respectively. The lower values of this metric indicate more similarity.

\subsection{Network data}
\label{sec:Network data}

\subsubsection{Synthetic datasets}

We utilize two types of synthetic networks:

\paragraph{LFR-benchmark:} As mentioned before, one of the most important features of real social networks is their community structure. To study the impact of this property on the performance of algorithms, we used the LFR-benchmark \cite{PhysRevE.80.016118,lancichinetti2008benchmark} which provides real networking charactristics, and its properties can be controlled to provide networks with build-in communities. The generated networks have $1000$ nodes with the characteristics shown in Table~\ref{tab:LFR benchmark characteristics}.
\begin{table}
\tbl{LFR-benchmark characteristics. (N: \#nodes, k: Avg degree, maxK: max degree, t1: minus exponent for the degree sequence, t2: minus exponent for the community size distribution, minC: Min community sizes, maxC: Max community sizes\label{tab:LFR benchmark characteristics}).}
{       
\begin{tabular}{|c|c|c|c|c|c|c|}
 \hline
\textrm{N}&
\textrm{k}&
\textrm{maxK}&
\textrm{t1}&
\textrm{t2}&
\textrm{minC}&
\textrm{maxC} \\ \hline
1000&15&50&2&1&20&50\\ 
\hline
\end{tabular}
}
\end{table}
An important attribute of these networks is a mixing parameter called $\mu$, which represents the network community structure. For each node, $\mu$ is a numeric value which is equal to the ratio of inter-community to entire links of that node. The value of $\mu$ can change between $0$ and $1$, and its alterations would change the community structure of the network. Larger values of $\mu$ corresponds to less community structure in the network, while decreasing this parameter would improve the community structure of the network. According to the experimental results, values over $0.5$ for $\mu$ are not suitable for highlighting the community structures. Therefore, we evaluated the competing methods against various types of LFR-benchmark by using different values of $\mu$ in the range of $[0.1, 0.2, 0.3, 0.4, 0.5]$.

\paragraph{Real structures:} We used the real co-authorship network of scientists dataset that inherently contains many community structures.

For each of these networks, we generated various artificial cascades by using the SNAP library \cite{SNAP} with different models, and utilized the average behaviors against different cascade models in our analysis to show independence of the proposed method from cascade models.

\subsubsection{Real datasets}

In this paper, we refer to a dataset as real, when its topology and corresponding diffusion are accessible.  We utilized two real networks over different months for LinkedIn and News of the World sites; with community structures, nodes and links that change overtime. All the specifications are listed in Table~\ref{tab:LinkdInNews}. 

\begin{table}
\tbl{\label{tab:LinkdInNews}
Real networks and diffusion information over them. The specifications are for different months. Each network had been changed during the months in a year \cite{conf/wsdm/Gomez-RodriguezLS13}.}
{
\includegraphics[scale=0.5, angle=270]{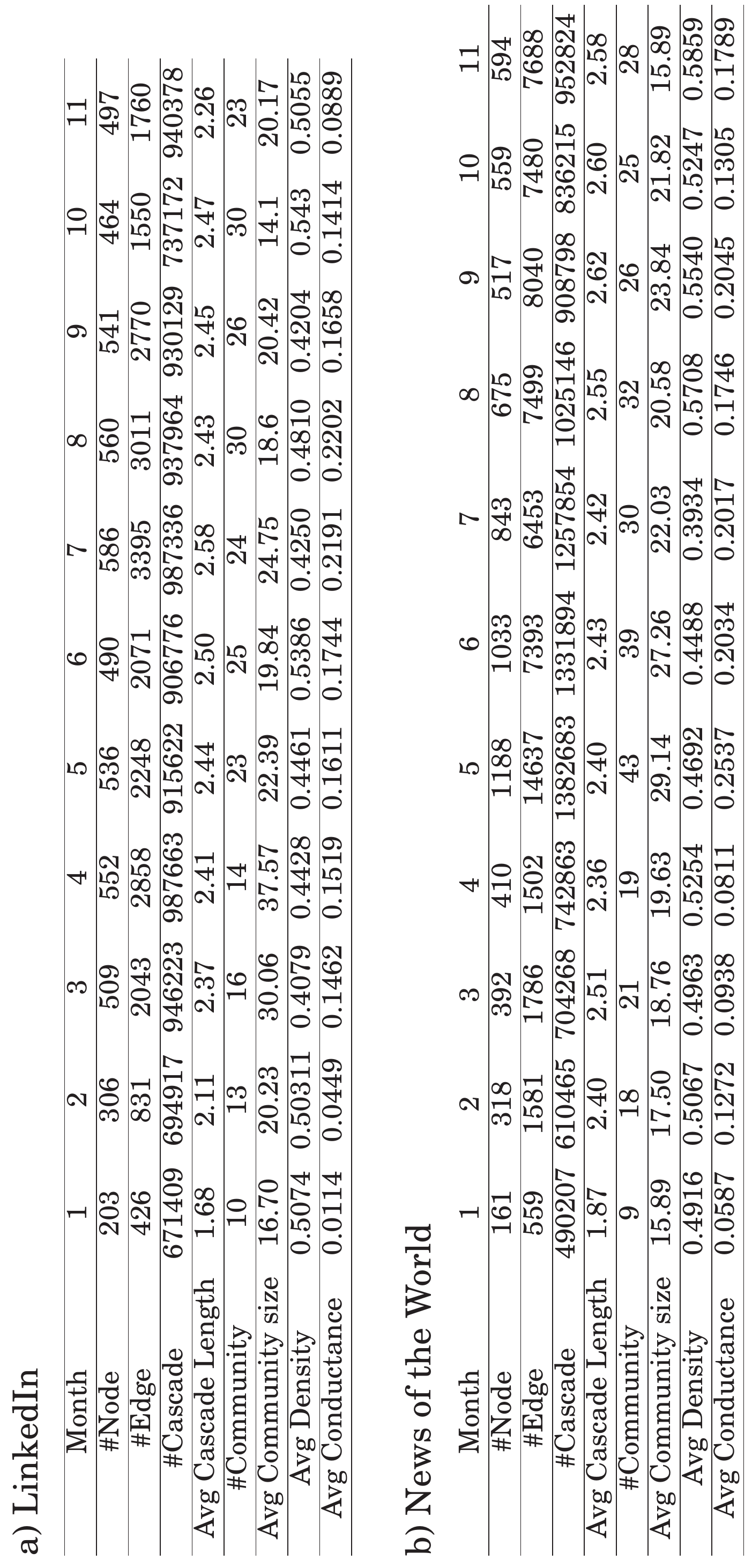}
}
\end{table}

\begin{figure}[]
\begin{center}
\includegraphics[width=0.32\textwidth]{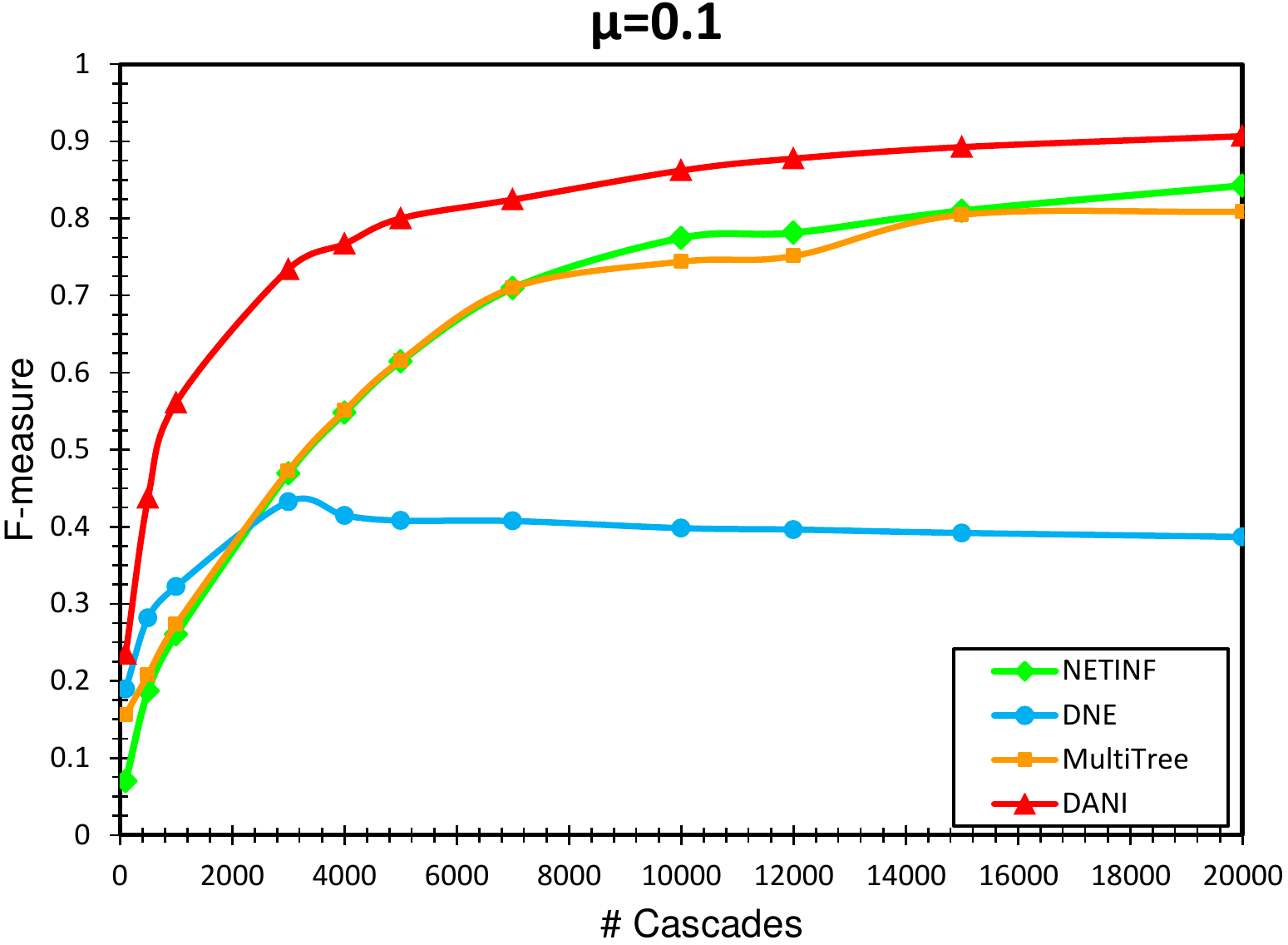}
\includegraphics[width=0.32\textwidth]{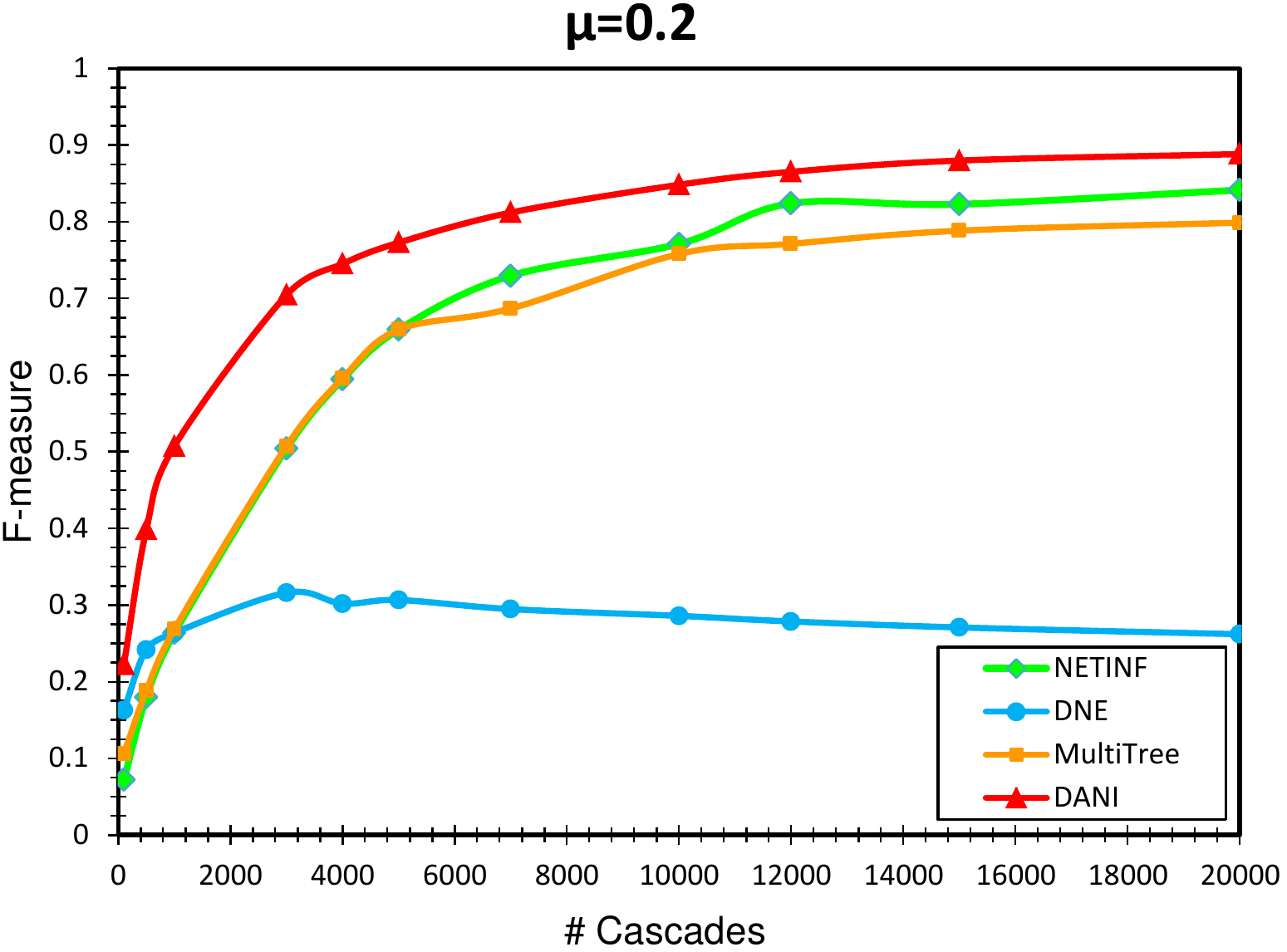}
\includegraphics[width=0.32\textwidth]{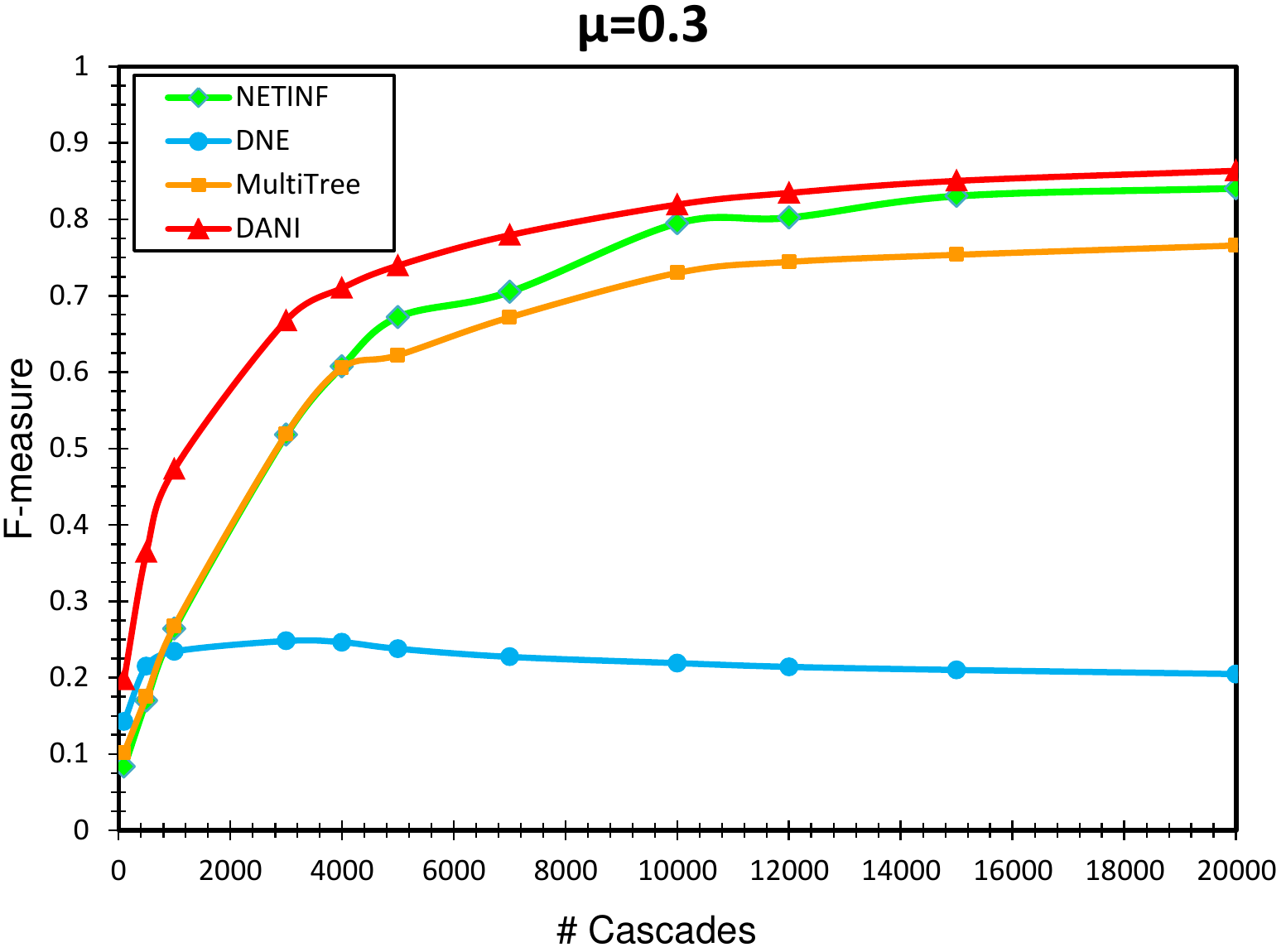}\\
\includegraphics[width=0.32\textwidth]{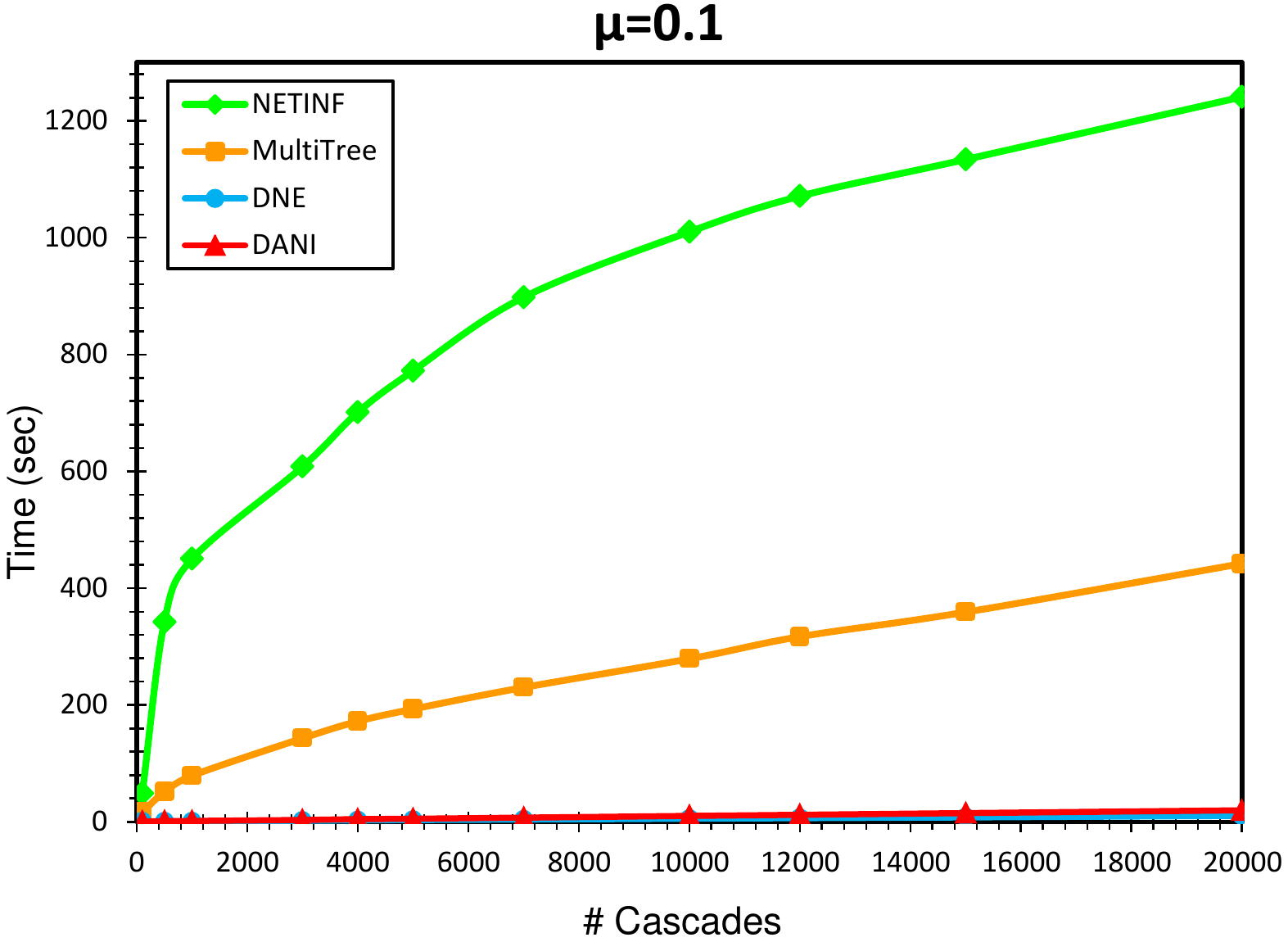}
\includegraphics[width=0.32\textwidth]{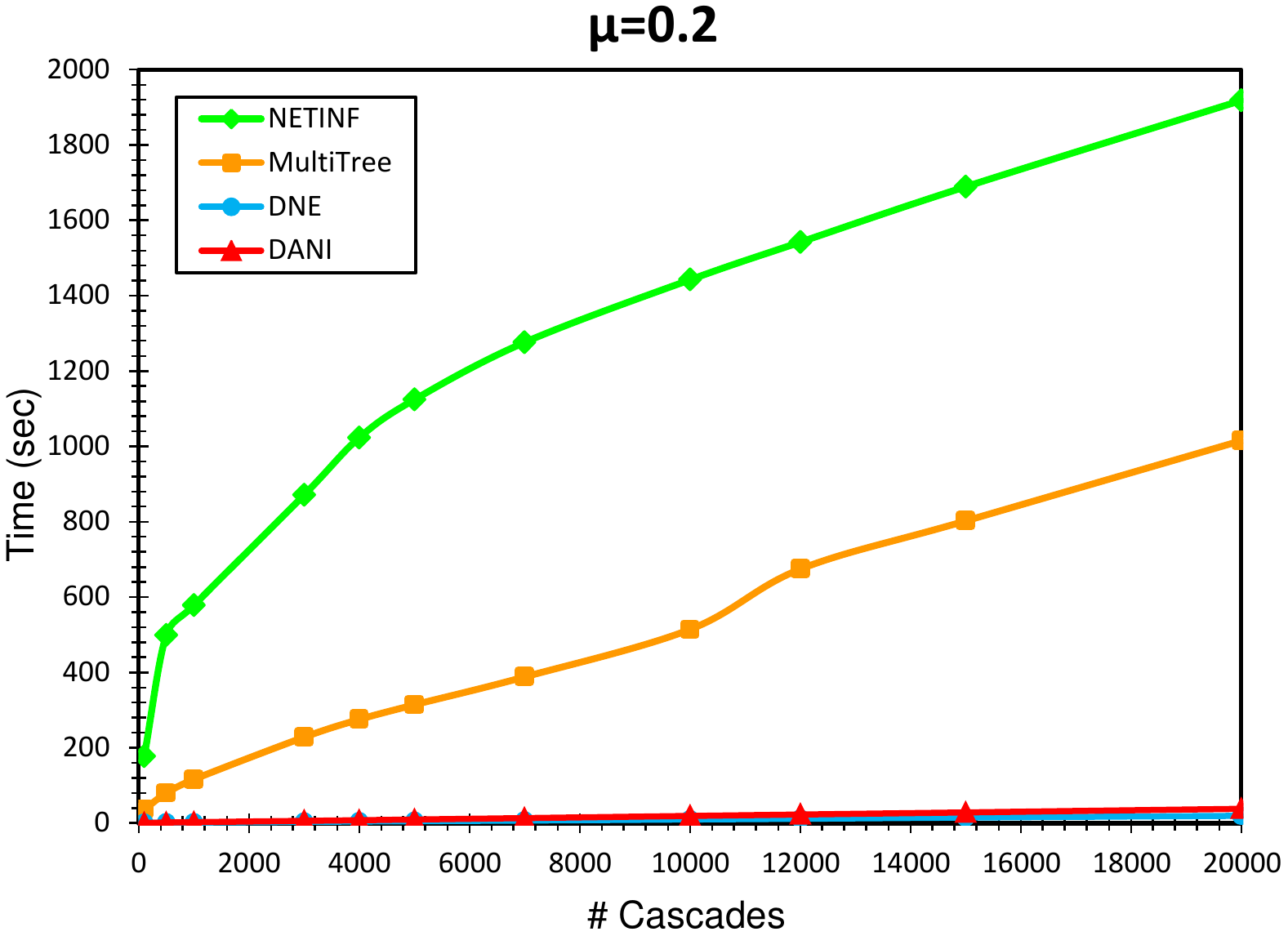}
\includegraphics[width=0.32\textwidth]{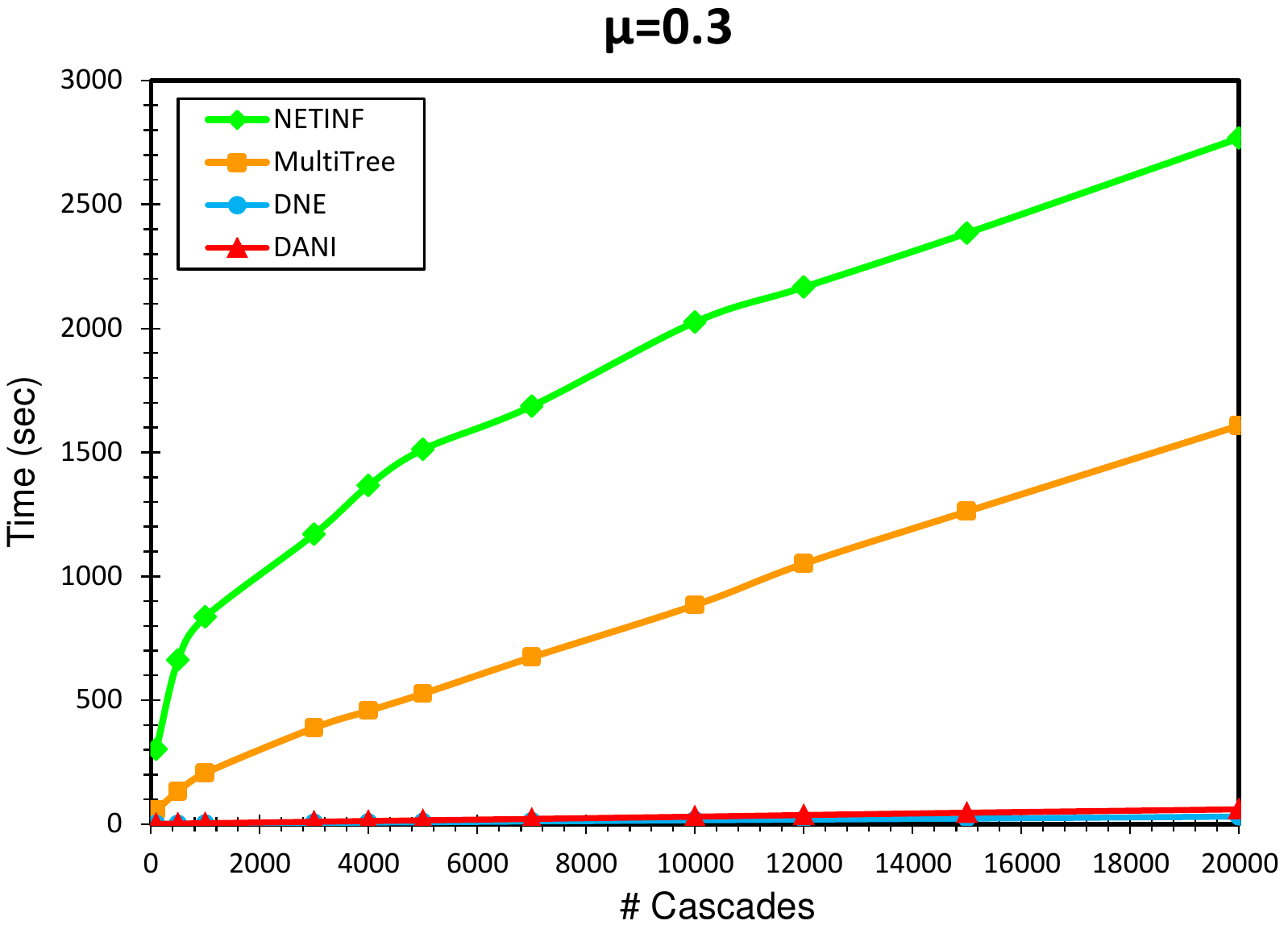}\\
\includegraphics[width=0.32\textwidth]{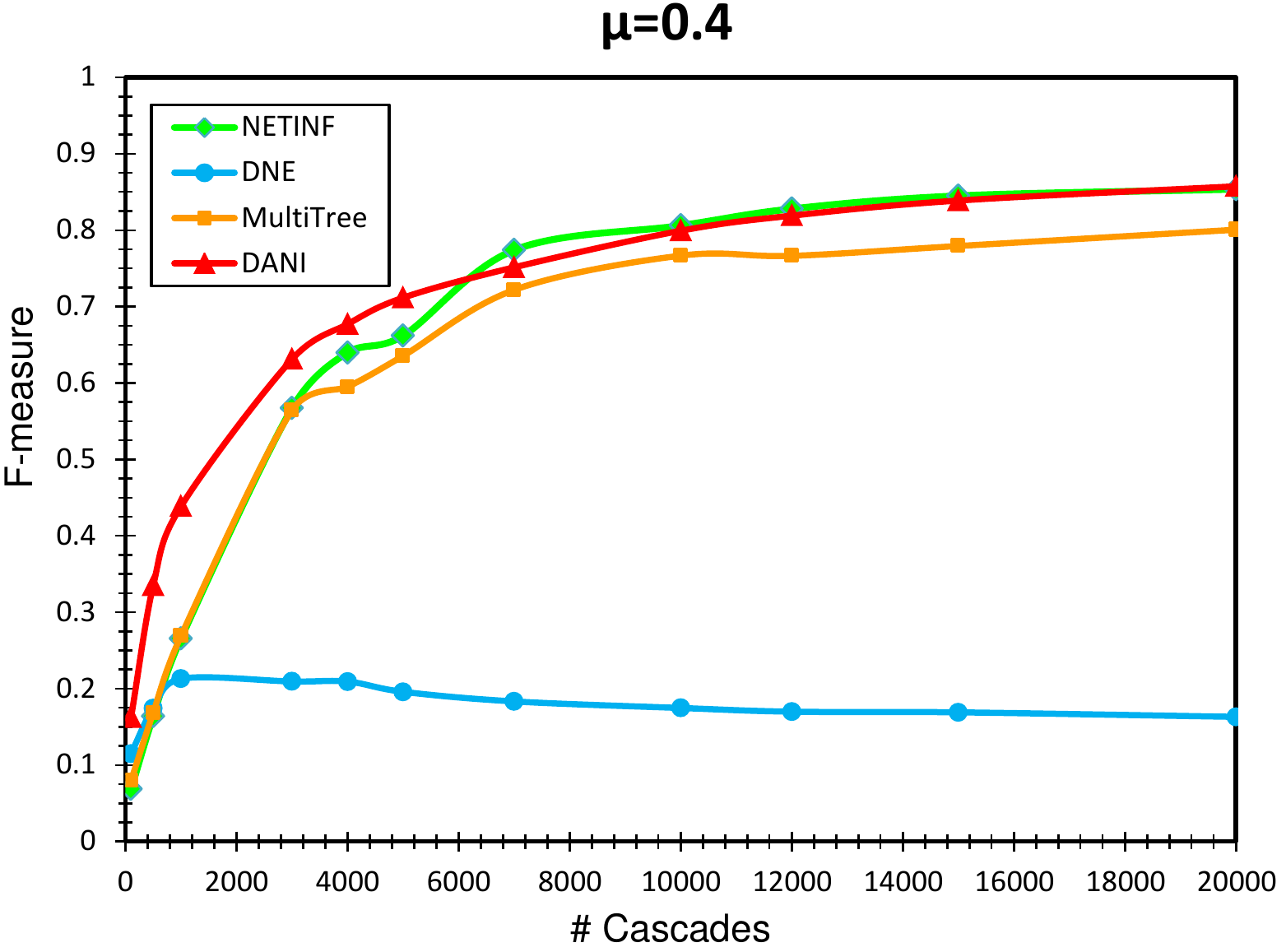}
\includegraphics[width=0.32\textwidth]{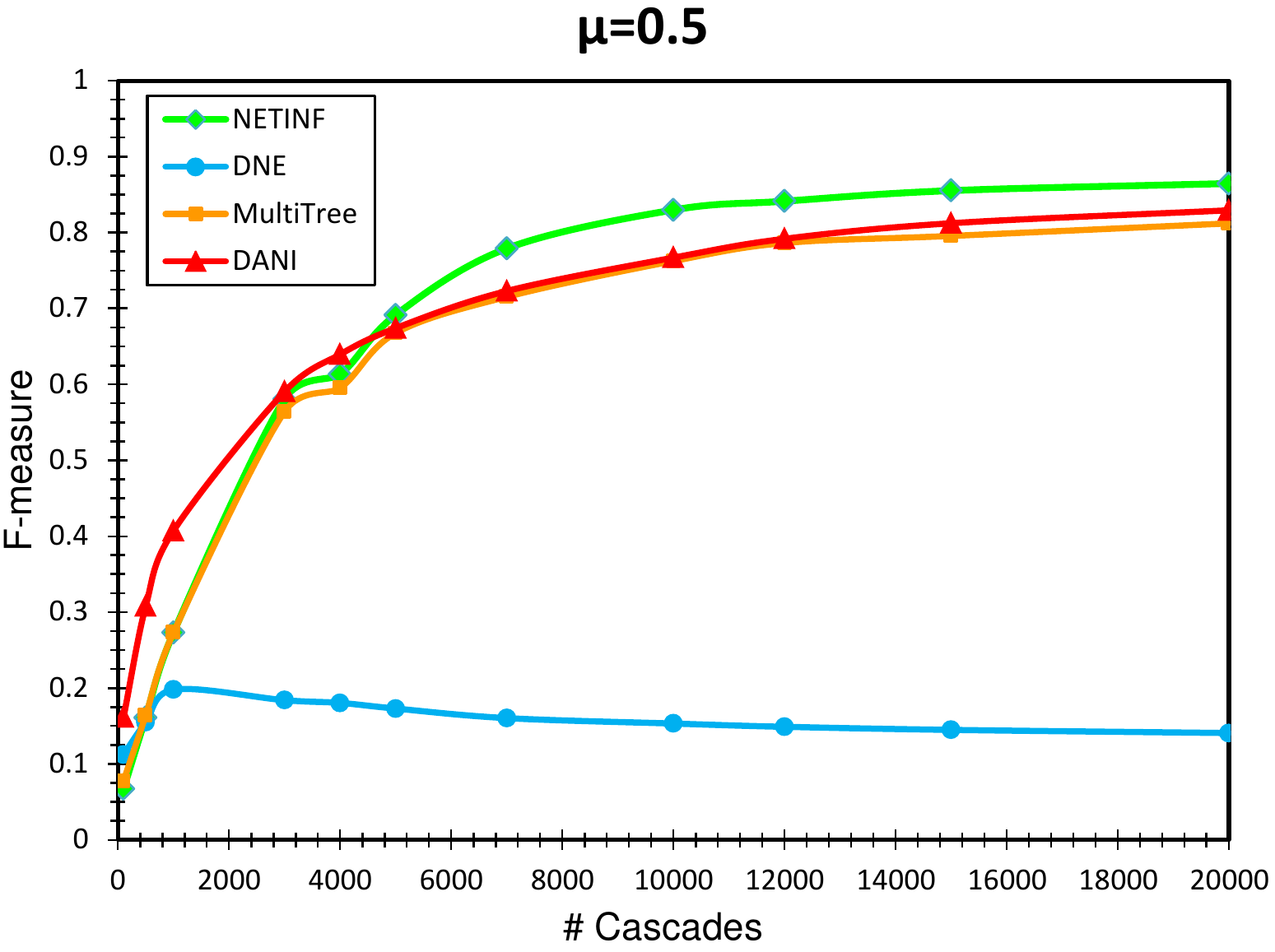}\\
\includegraphics[width=0.32\textwidth]{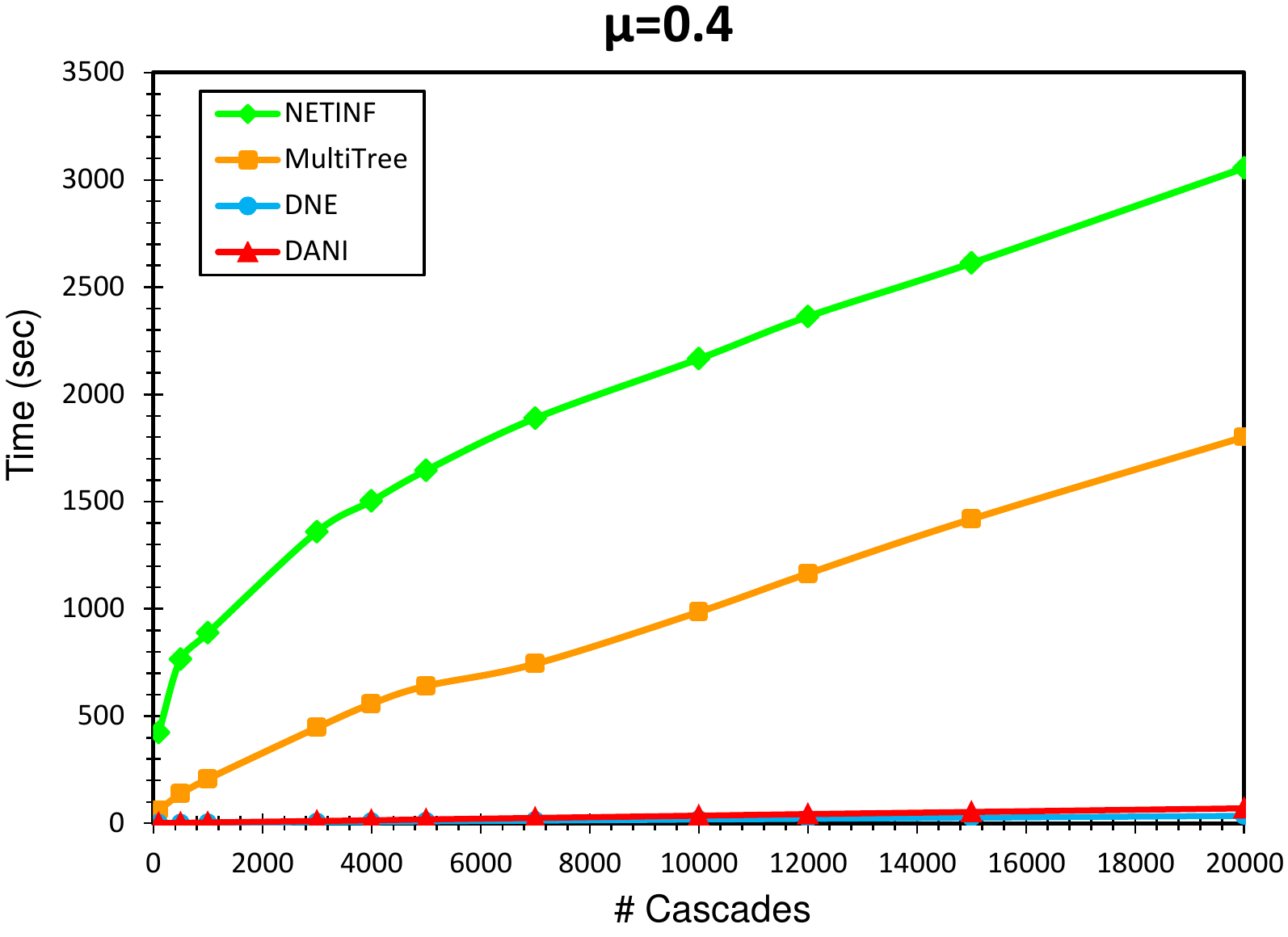}
\includegraphics[width=0.32\textwidth]{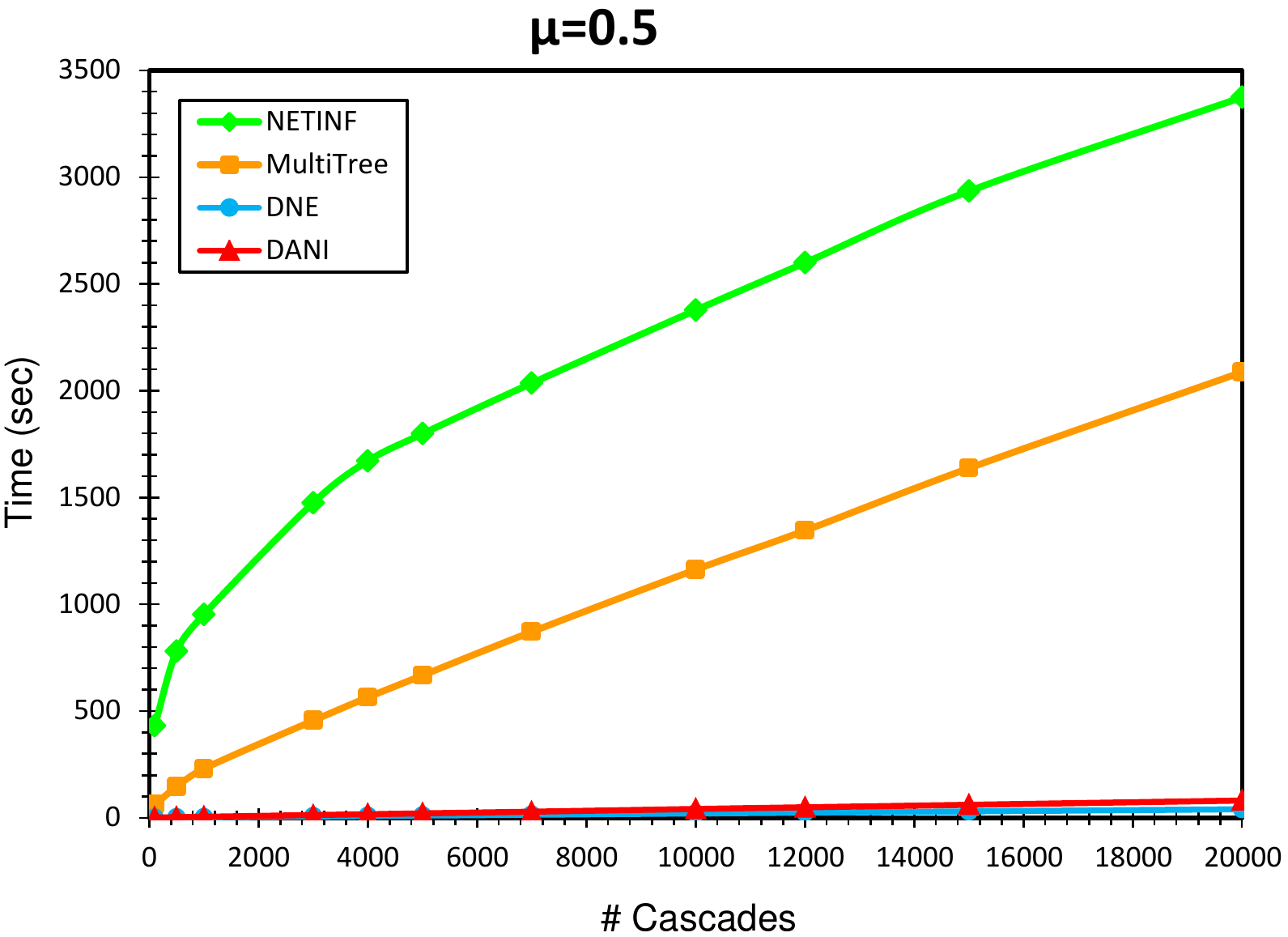}
\caption{\label{fig:LFRFMeasure} (Color online) F-measure and running time against different number of cascades in LFR-benchmark network. Each part is for a special mixing parameter ($\mu$). Figures show averages over 10 runs.}
\end{center}
\end{figure}

These datasets have been gathered by using the MemeTracker from March 2011 to February 2012 \cite{MemeTracker}. We split all the months into eleven intervals and discuss the performance of algorithms in these intervals. The network visualizations available at \cite{conf/wsdm/Gomez-RodriguezLS13,INFOPATHwebsite} maybe used to demonstrate the changes in structures and illustrate the communities for different months. 

\subsection{Results}
\label{sec:Results}

We analyzed the results of three types of qualified networks based on the evaluation metrics described above.  The interpretation of the results are provided in the following subsections.

\subsubsection{Network inference:}

\paragraph{\textbf{F-measure}:} As Fig.~\ref{fig:LFRFMeasure} illustrates, the proposed algorithm (DANI) has a better performance in inferring underlying networks compared to the competing methods. As shown, DANI has a higher F-measure for LFR-benchmark networks with different community structures (different values of $\mu$). In this figure, X-axis represents the number of cascades in the network. For various community structures (mixing parameters) and number of cascades, DANI outperformed DNE and MultiTree.

When $\mu$ changes from $0.1$ to $0.3$, DANI achieves higher F-measures in lower running time than NETINF. As the community structure fades away ($\mu\succ0.3$), F-measure for DANI is four times better than DNE, and two times better than MultiTree. When the number of cascades over the network becomes more than the number of links ($\succ7692$), F-measure for NETINF is $0.1$ more than the F-measure for DANI, while the time complexity of NETINF was much more than DANI. For the same case, MultiTree achieved the same F-measure as NETINF, while its running time was better than NETINF, but higher than DANI.

\begin{figure}[]
\begin{center}
\includegraphics[width=0.32\textwidth]{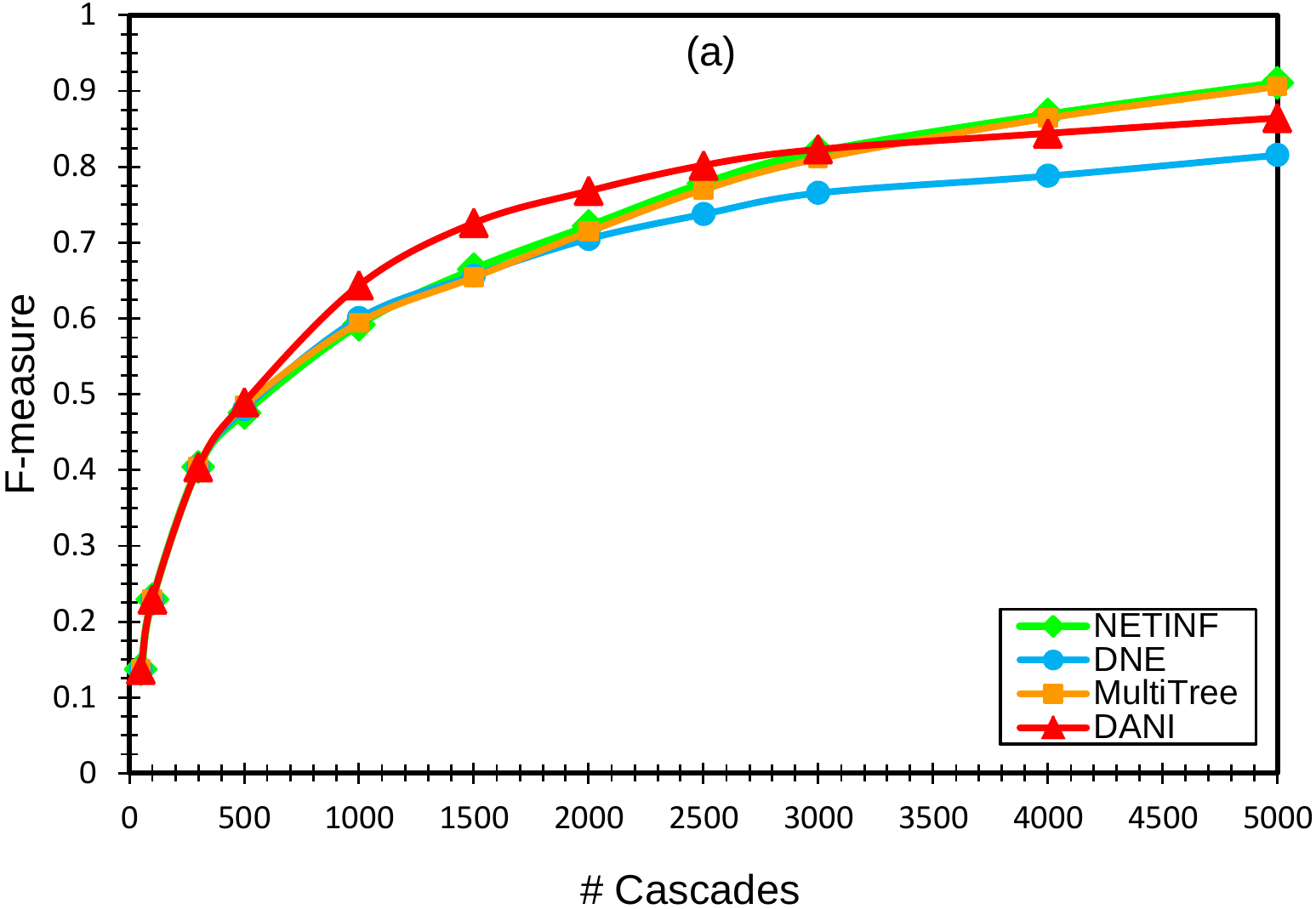}
\includegraphics[width=0.32\textwidth]{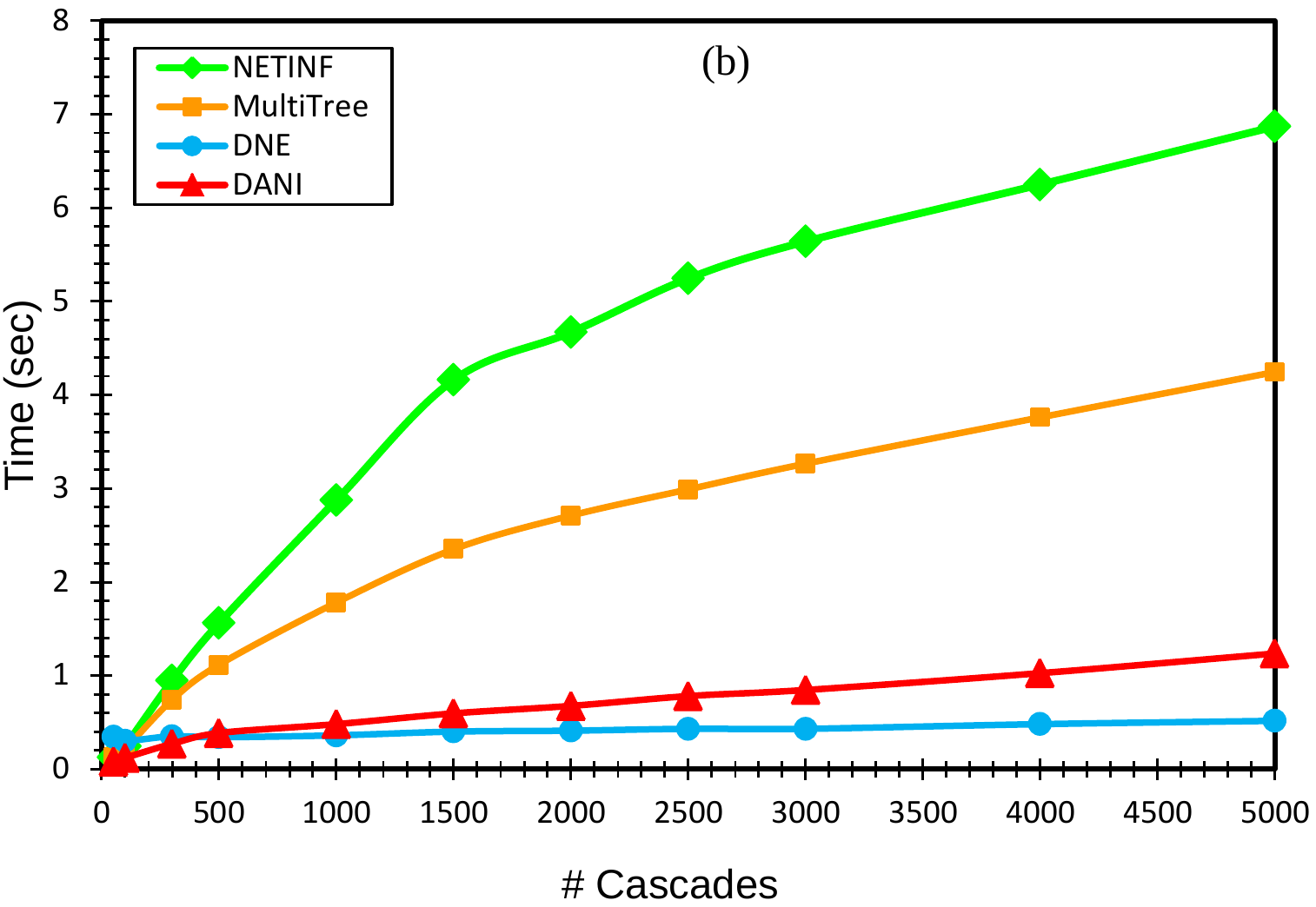}
\includegraphics[width=0.32\textwidth]{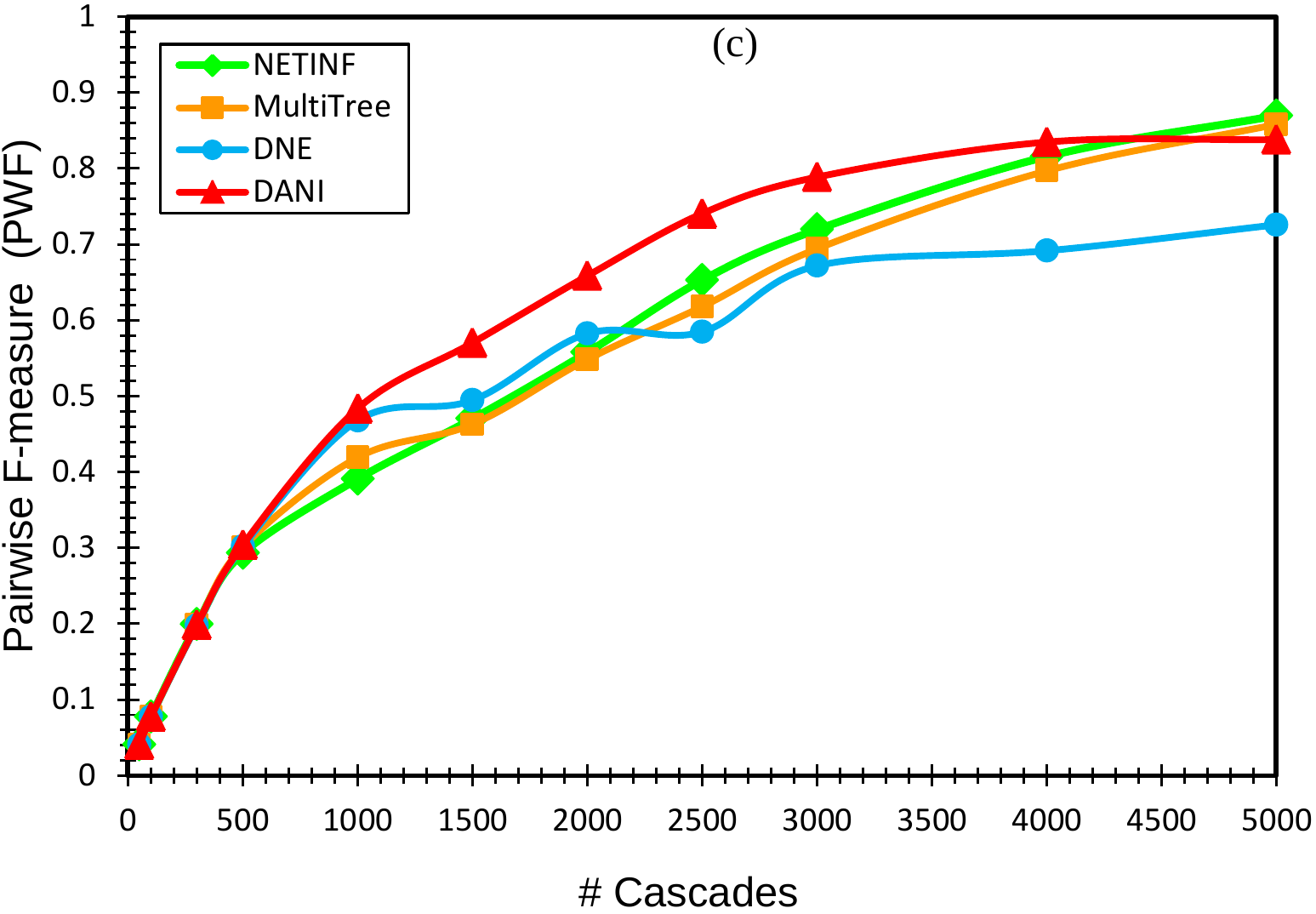}\\
\includegraphics[width=0.8\textwidth]{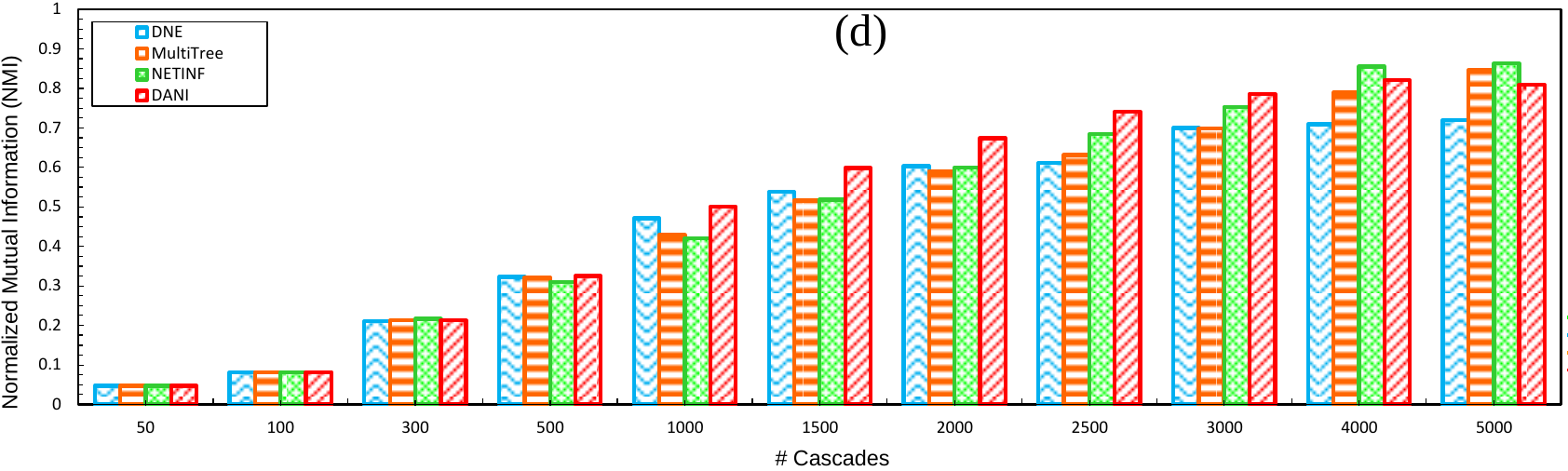}
\caption{\label{fig:NetScience}(Color online) Test of the algorithms on a real network with the NetScience \cite{newman2006finding} artificial cascade; $1590$ nodes, $2742$ edges, and $191$ communities. The communities are detected by the OSLOM algorithm \cite{lancichinetti2011finding}. Results are shown for the inferring scope: a) F-measure, and b) Running Time, and the community detection scope: c) Pairwise F-measure, and d) Normalized Mutual Information. Figures show averages over 10 runs.}
\end{center}
\end{figure}

For the NetScience dataset (Fig.~\ref{fig:NetScience}), F-measure of DANI was consistently more than DNE. For small number of cascades, DANI had higher F-measure than NETINF and MultiTree. Therefore, when the number of cascades are higher than the number of the links ($\succ2742$), NETINF and MultiTree infer the links accurately, but they require considerably higher running time compared to DNE and DANI.

\begin{figure}[]

\includegraphics[width=0.5\textwidth]{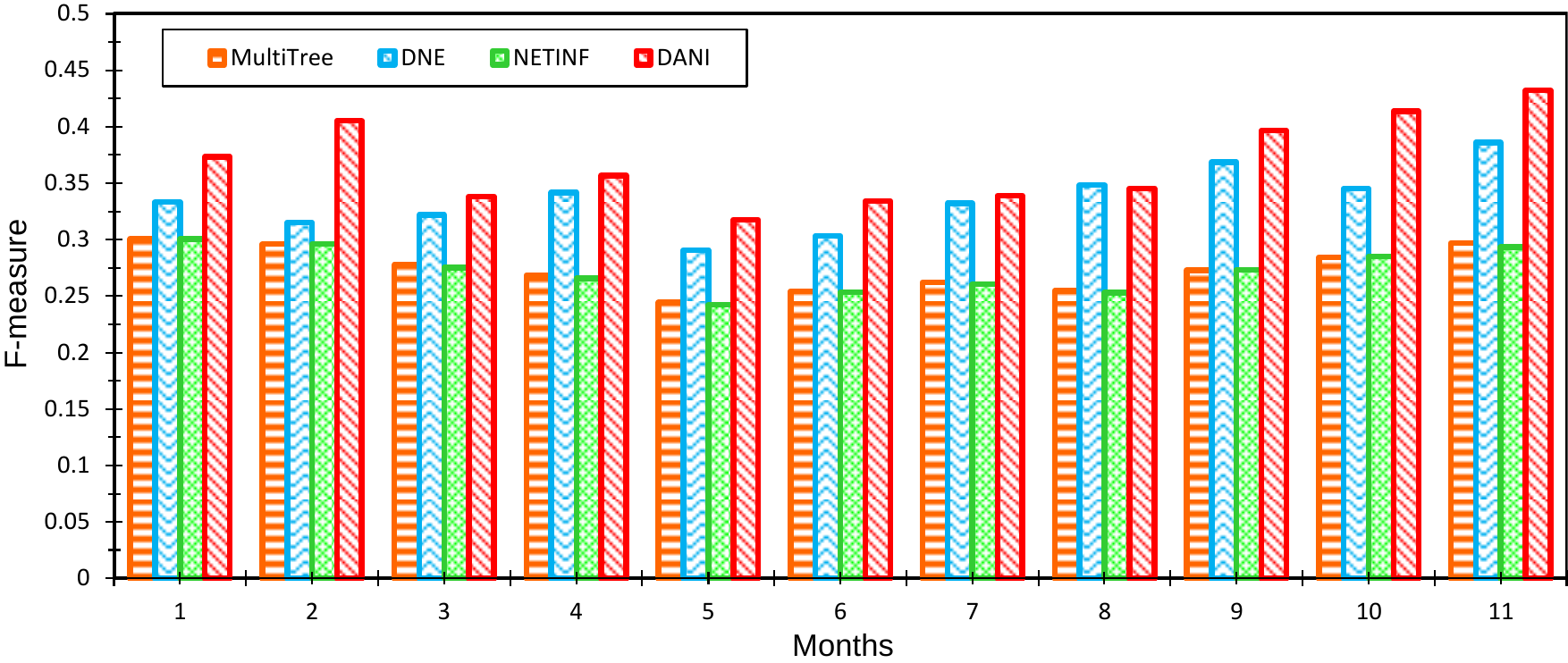}
\includegraphics[width=0.5\textwidth]{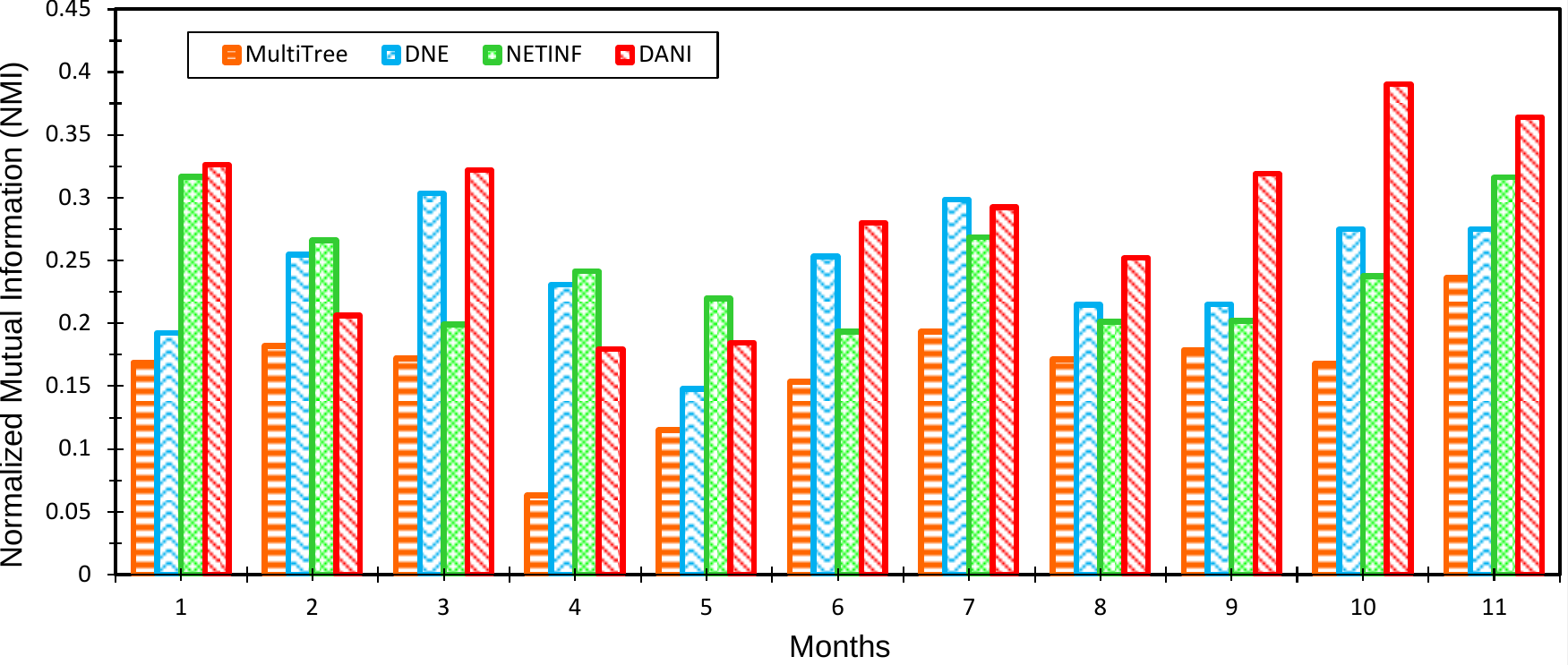}\\
\begin{center}
\includegraphics[width=0.5\textwidth]{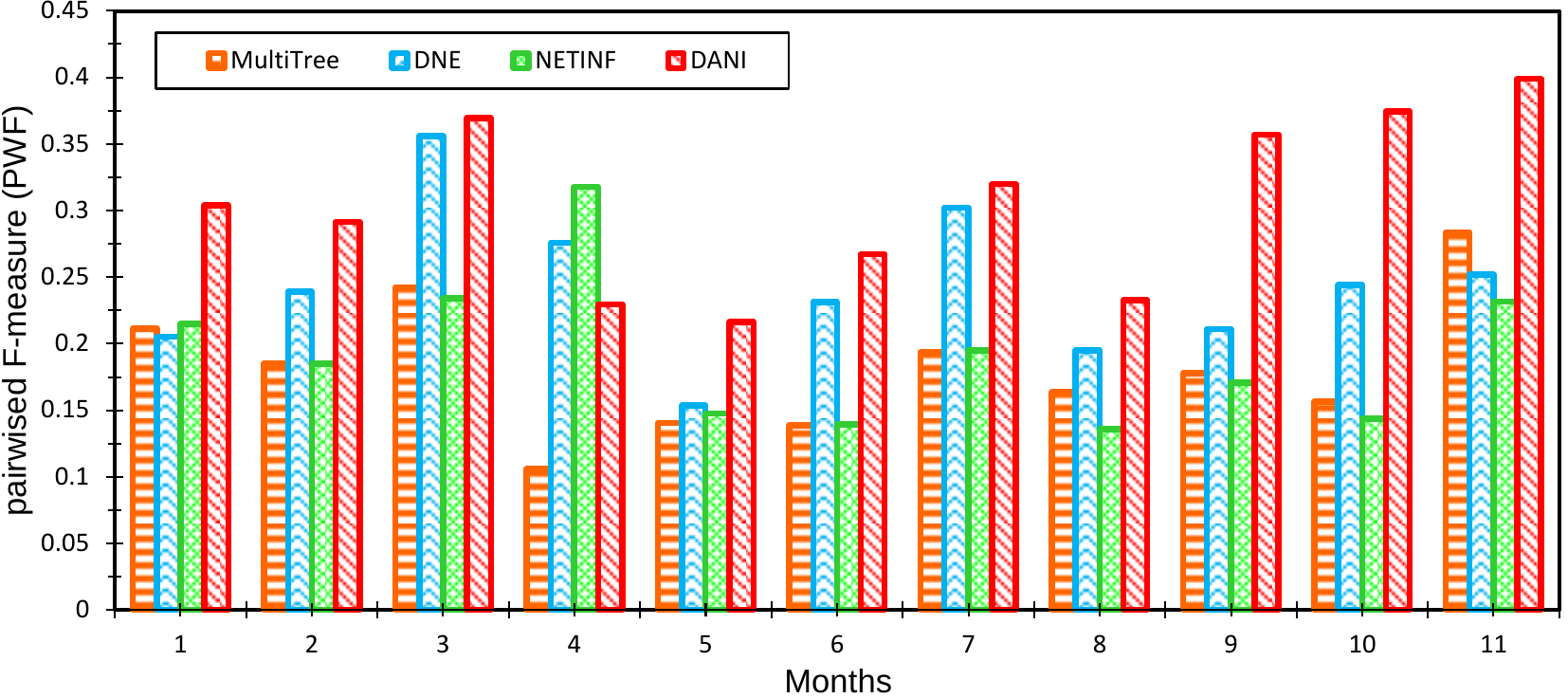}
\includegraphics[width=0.4\textwidth]{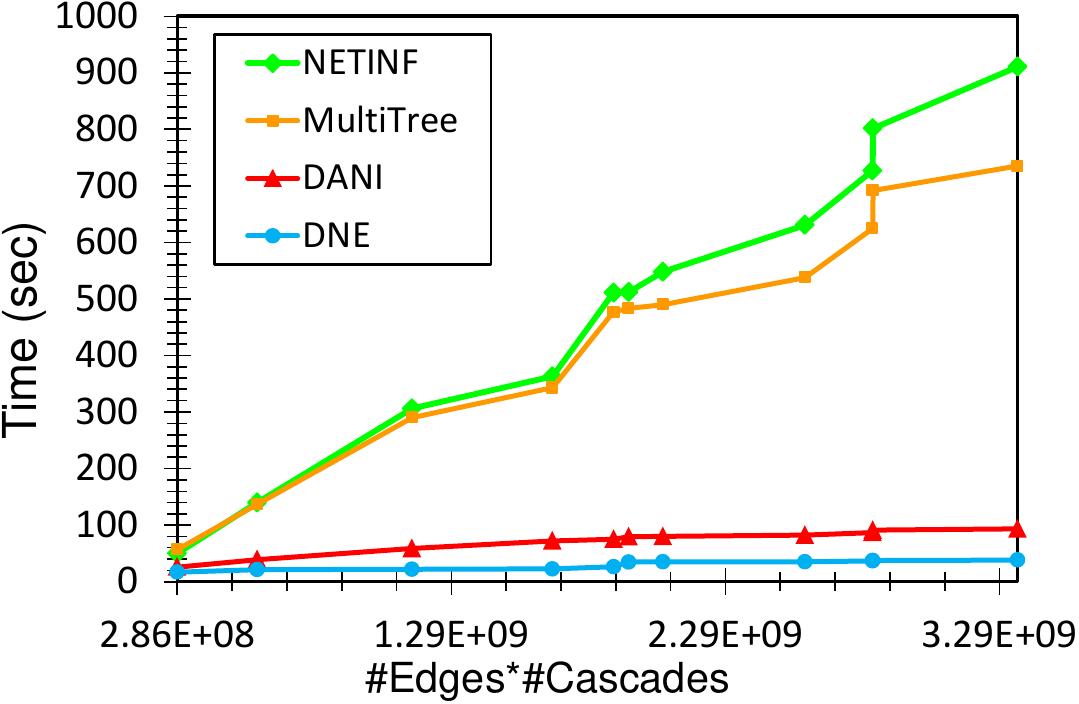}
\end{center}
\caption{\label{fig:LinkedInPlot}(Color online) Tests of the algorithms on LinkedIn dataset in $11$ consecutive interval months.}

\end{figure}

\begin{figure}[]
\begin{center}
\includegraphics[width=0.46\textwidth]{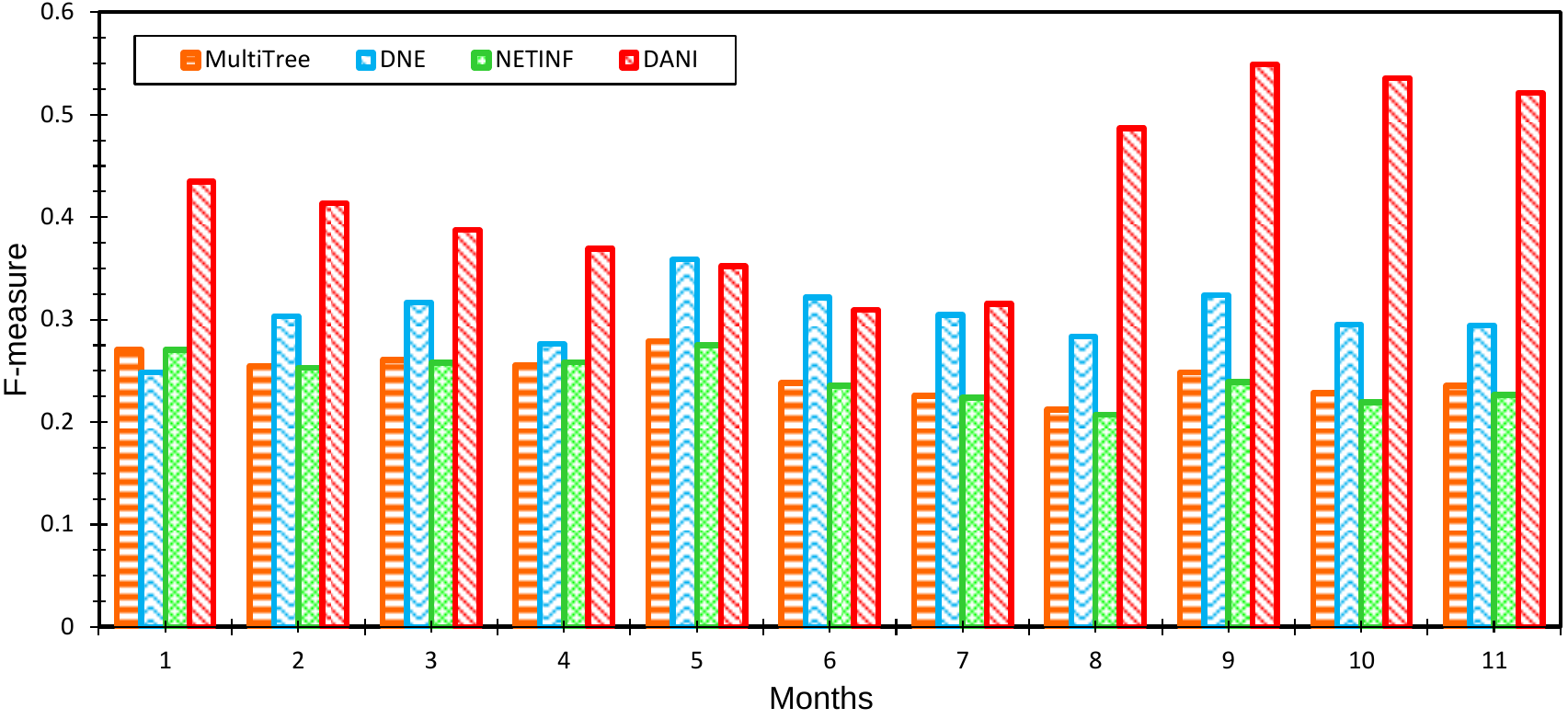}
\includegraphics[width=0.5\textwidth]{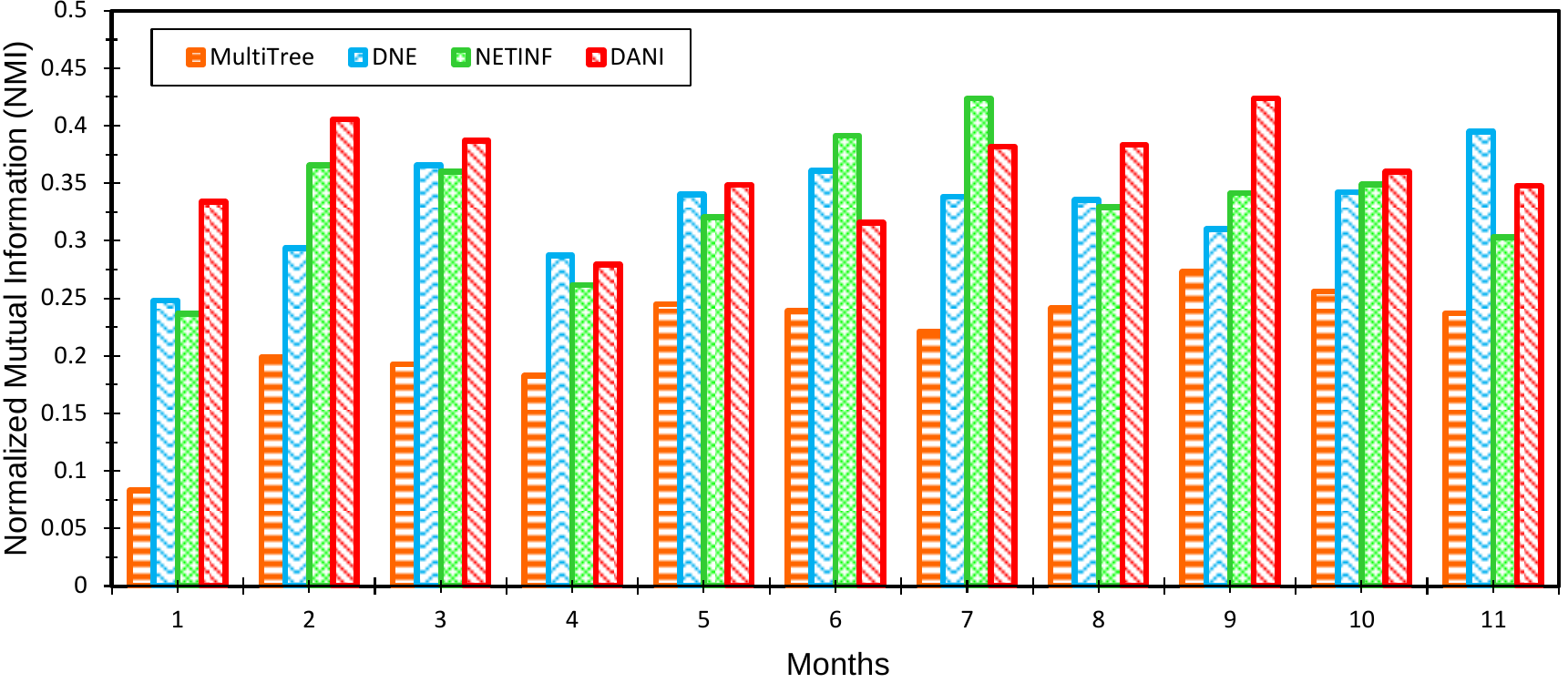}\\
\includegraphics[width=0.5\textwidth]{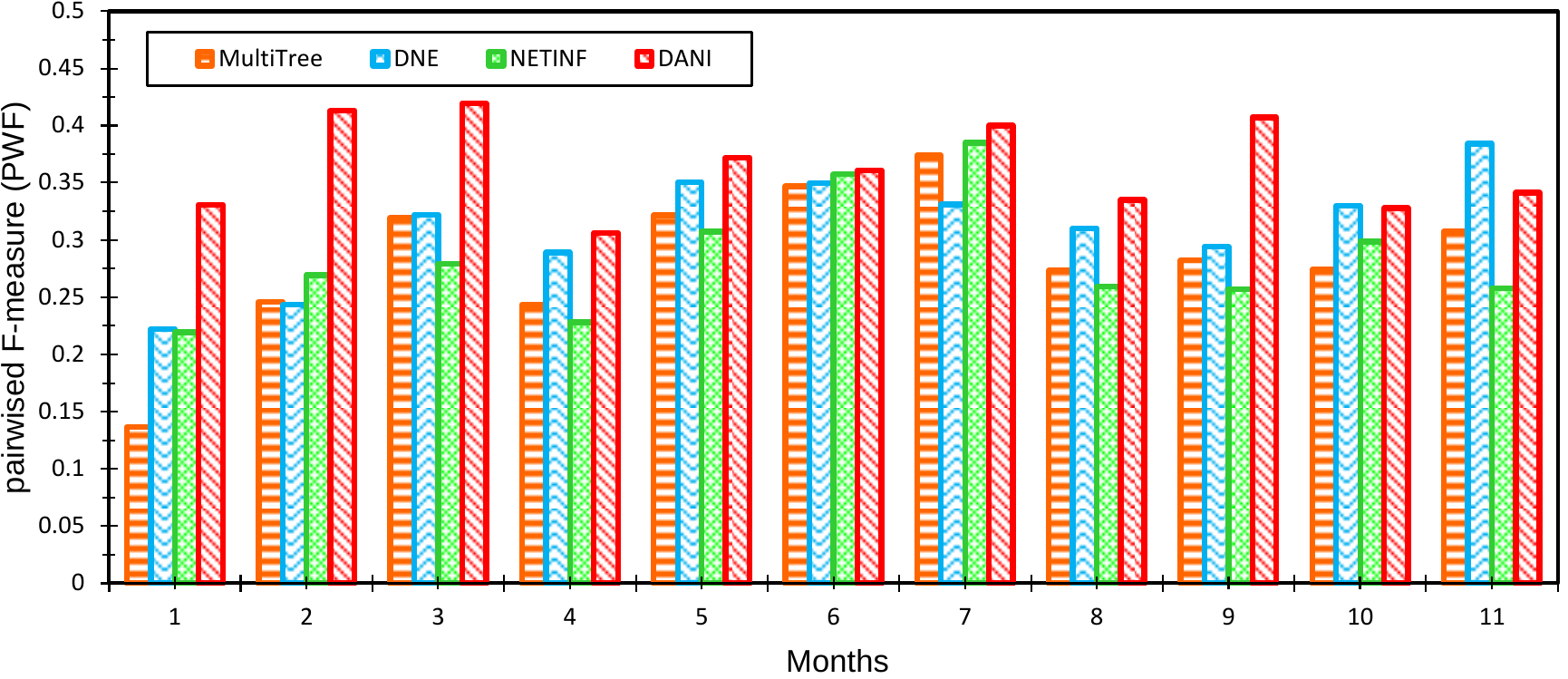}
\includegraphics[width=0.4\textwidth]{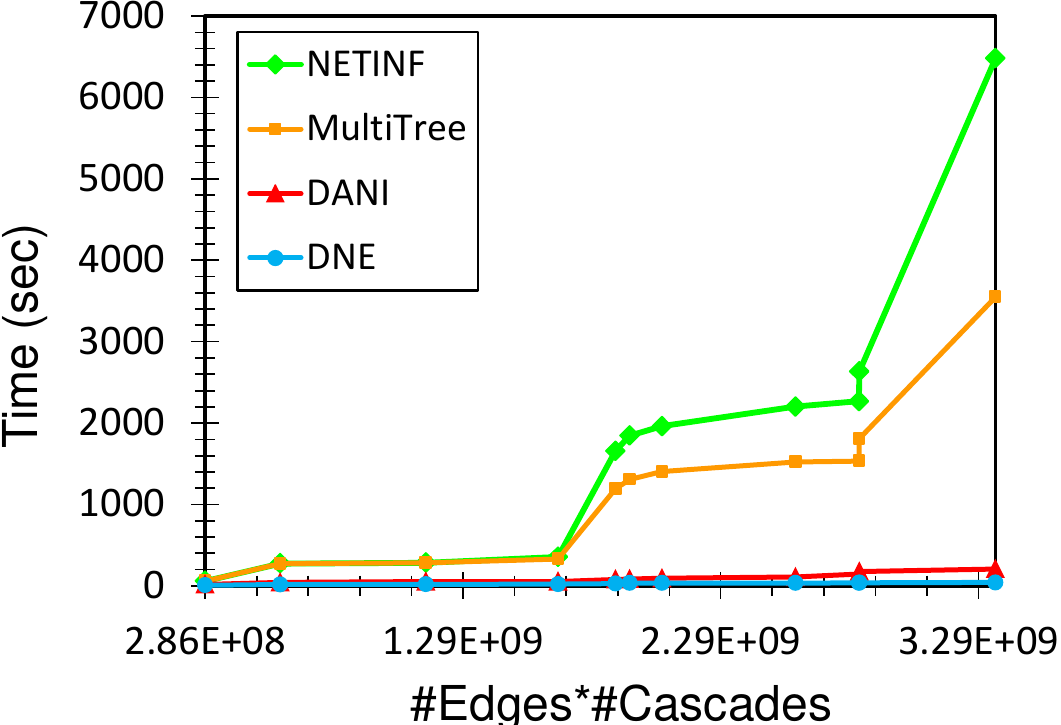}
\caption{\label{fig:NewsoftheWorldPlot}(Color online) Tests of the algorithms on News of the World dataset in $11$ consecutive interval months.}

\end{center}
\end{figure}

For the real datasets, we display the results for $11$ different months (Figs.~\ref{fig:LinkedInPlot} and ~\ref{fig:NewsoftheWorldPlot}). It can be seen that DANI with higher F-measure retrieves more network links. 


Investigations show that most of the cascades over real networks are short, and the number of long diffusion processes are small \cite{Leskovec:2005:pakdd}. A major contagion over networks usually does not have one seed user, and broadcasts in several different locations on the network. Each cascade motion from different users is being seen as a separated cascade. Furthermore, the wide spread of a contagion may situate on a network, but inevitably does not occur during a long cascade. 

\begin{figure}[]
\begin{center}
\centering
\includegraphics[width=0.4\textwidth]{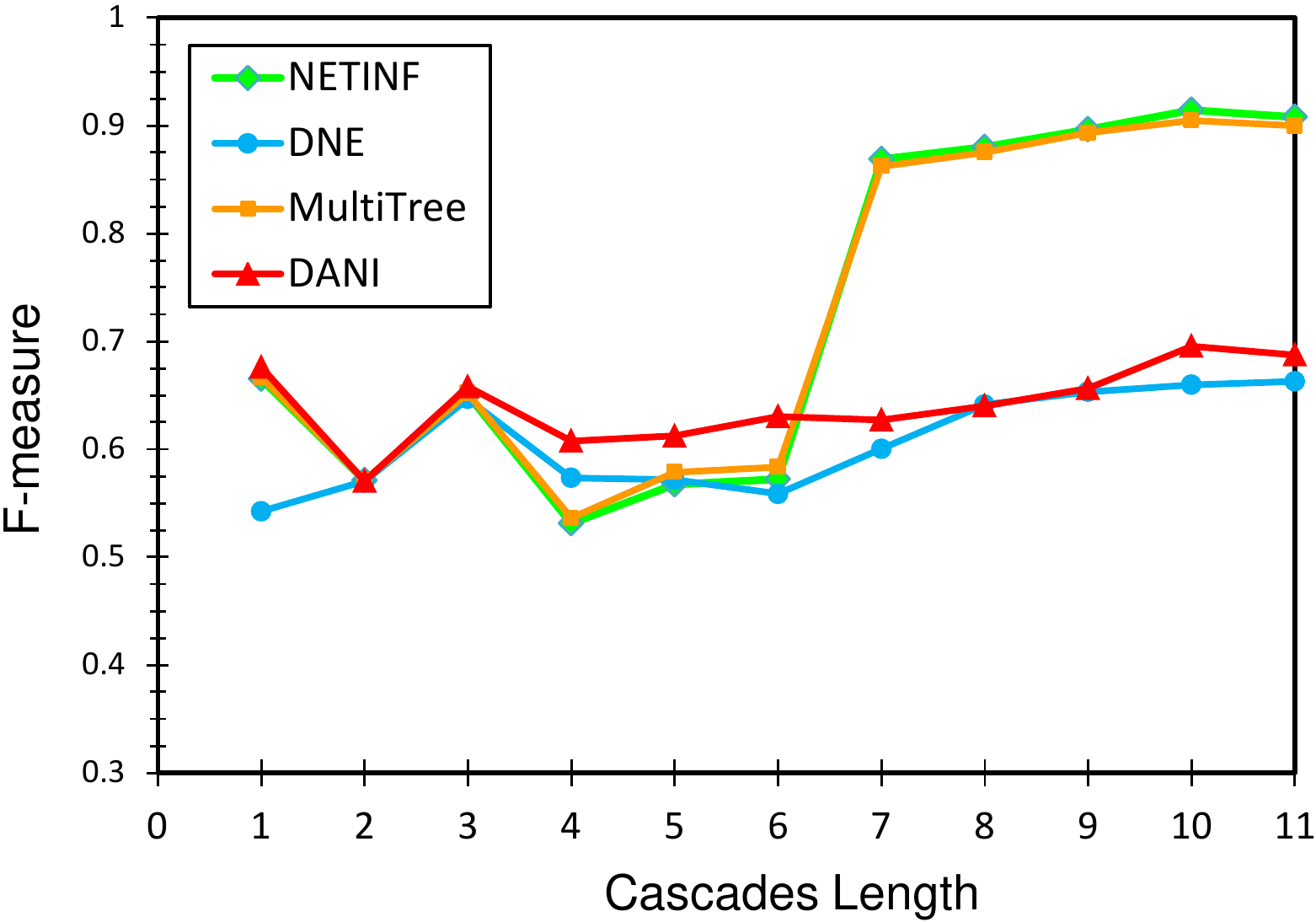}
\caption{\label{fig:LengthTest}}(Color online) The F-measure across different cascade lengths (in $350$ cascades) for the Football dataset \cite{Girvan2002}, with $115$ nodes and $615$ links.
\end{center}
\end{figure}

By checking the number and the length of cascades in these datasets, we concluded that the lengths of cascades are short in every month (Table~\ref{tab:LinkdInNews}). Based on our experiments on a real network with artificial set of cascades (Fig.~\ref{fig:LengthTest}), NETINF and MultiTree accuracy are low in short cascade lengths, and their dependency on the length of cascades is greater than DANI. However, MultiTree serves better than NETINF in short cascades, and therefore performs better in real networks.

On average, in LinkedIn network, DANI has $35.1\%$ improvement in F-measure versus NETINF, $34.3\%$ versus MultiTree, and $10.0\%$ versus DNE. Moreover, in News of the World network, this improvement is $77.91\%$ against NETINF, $74.40\%$ against MultiTree, and $42.26\%$ against DNE.

\paragraph{\textbf{Number of nodes}:} Table~\ref{tab:Behavior} displays the behavior of these algorithms for the number of detected nodes of the inferred networks. DNE can just retrieve a few set of nodes, and most of the links of the networks are not inferred. Although, NETINF and MultiTree are better than DNE in this regard, but their loss error is more than $20\%$. DANI infers the nodes with an accuracy close to $90\%$. In other words, DANI infers maximum number of nodes compare to the other methods. 
The results show that DANI is more capable in detecting nodes in the inference process, which has not been considered in prior works.

\paragraph{\textbf{Other network properties}:} The relative error of network properties between the inferred and underlying network is illustrated in Table~\ref{tab:Behavior}. In most cases, DANI has much less error than baseline methods in term of nodes degree. In addition, the clustering coefficient metric is better maintained by DANI.

\paragraph{\textbf{Running time}:} In order to have a fair comparison, we used the available C++ implementations in SNAP library \cite{SNAP} for all prior algorithms, and implemented DANI in the same framework. We compared runtimes in the LFR-benchmark (Fig.~\ref{fig:LFRFMeasure}) that showed DANI was several order of magnitude faster. In the NetScience dataset for different number of cascades, the run time is shown in Fig.~\ref{fig:NetScience}. For running time in real data sets, similar to the approach used in \cite{Fastinf11}, we measured time versus number of cascades and links as shown in Figs.~\ref{fig:LinkedInPlot} and ~\ref{fig:NewsoftheWorldPlot}.  On average, DANI is $85.35\%$ faster than NETINF, and $84.11\%$ faster than Multitree. In general, the running time of algorithms is ordered as ${t_{NETINF}} \succ {t_{MultiTree}} \succ {t_{DANI}} \succ {t_{DNE}}$. 

In all cases, our method has lower run time than tree-based algorithms, and its speed is close to DNE. The running time performance must be analyzed against inference accuracy and preserving community structures. 

As illustrated, DANI achieves acceptable results with low running times, while other algorithms require higher running times to infer the networks with high accuracy. 

\paragraph{\textbf{Evaluating the scalability}:}  As we already mentioned the time complexity of DANI depends on the number of users involved in cascades (nodes of network) which could be extremely large. Here, we demonstrate the scalability of DANI against the competing methods by showing the results on a real large scale network. In addition, parallel methods such as MapReduce can be used for making DANI even more scalable, but to observe fairness, we did not utilized any of them here. We studied a large Twitter network that sampled public tweets from the Twitter streaming API in the range of 24 March to 25 April at 2012 \cite{weng2013virality}. By utilizing the follower network and hashtag retweeting, we did some preprocessing by: 1) Considering any unique pair of hashtag and source ID as a separate cascade, 2) Keeping links that were being involved in at least one cascade, 3) Choosing users taking part in more than 2 cascades, and 4) Removing nodes of cascade that were not a member of network nodes. The result was a network of 124023 nodes, 182783 links, and 213875 cascades with an average length of 2.29. While DANI correctly inferred 79.64\% of the links in terms of F-measure during 1 hour and 4 minutes, the other three competing methods did not yield any reasonable results during this time period.

\subsubsection{Preserving community structure:}
One of the important properties of social networks is their community structure and our goal in this paper is to preserve this property during the network inference process. To achieve this goal, the inferred network must have the same community structure as the underlying network. Therefore, the properties such as density, conductance, community size and number of communities are some important characteristics, which we considered in our evaluations. 

To detect communities in real and inferred networks, we reviewed different community detection algorithms to find an algorithm which can work with a variety of networks such as directed and undirected, support overlapping communities, detect communities with high accuracy, not require prior knowledge about the number of output communities, and require low memory and time for a sparse graph. Finally, the OSLOM algorithm was chosen which meets the above requirements. OSLOM is the first algorithm that uses statistical features to identify the communities. To assess the accuracy, OSLOM uses local modularity measurement to avoid global errors \cite{lancichinetti2011finding}.

\begin{figure}[]
\begin{center}
\includegraphics[width=0.32\textwidth]{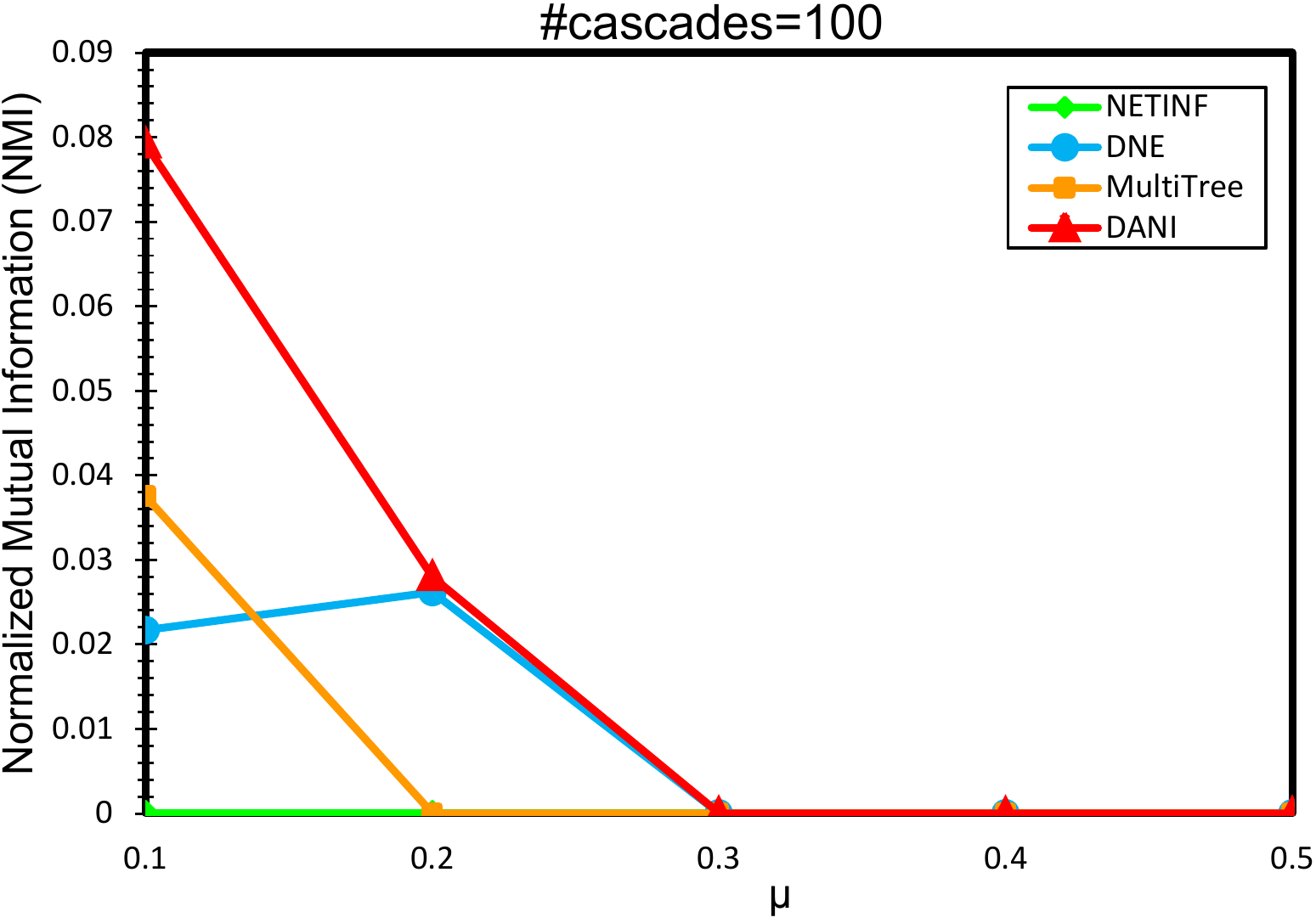}
\includegraphics[width=0.32\textwidth]{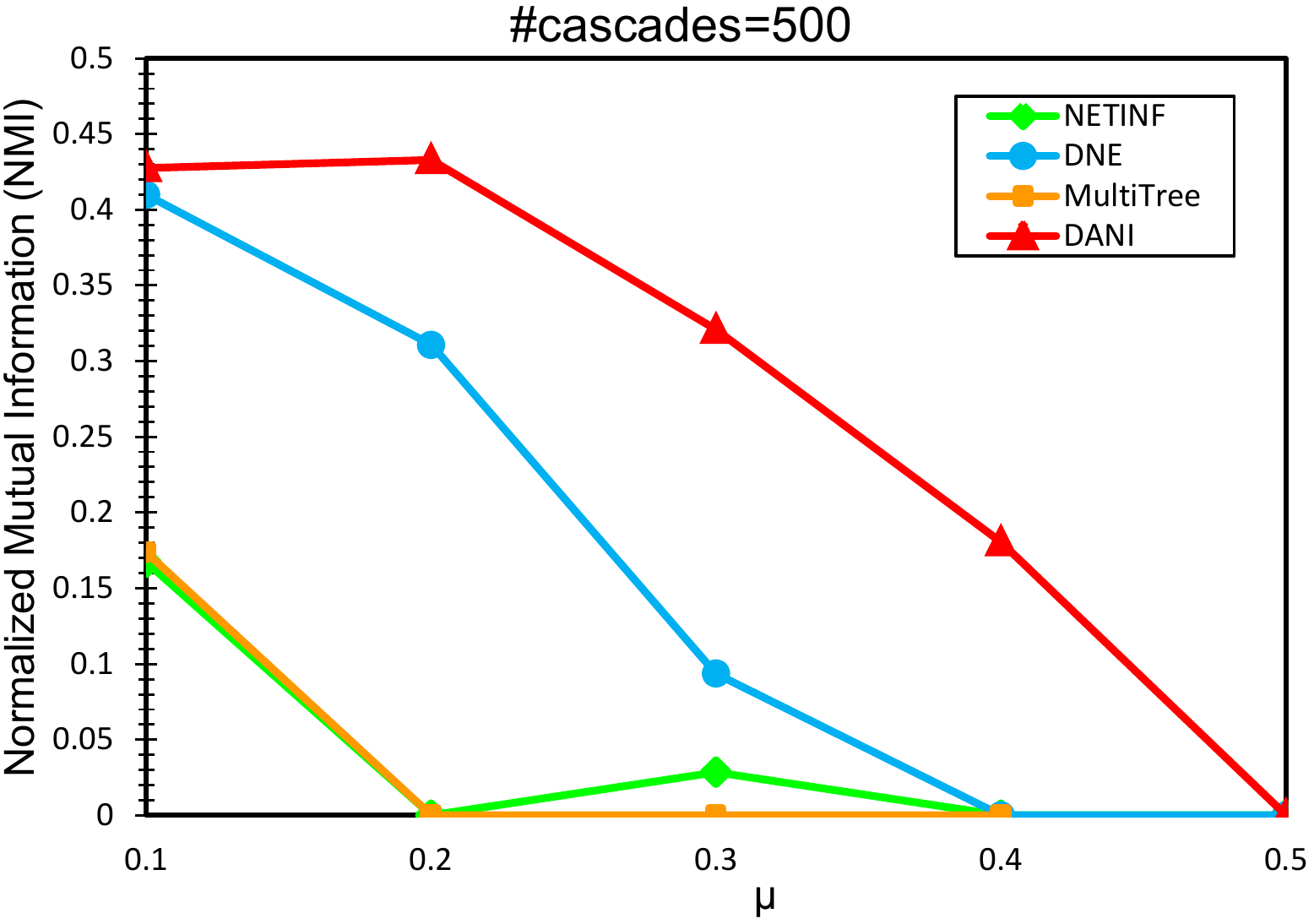}
\includegraphics[width=0.32\textwidth]{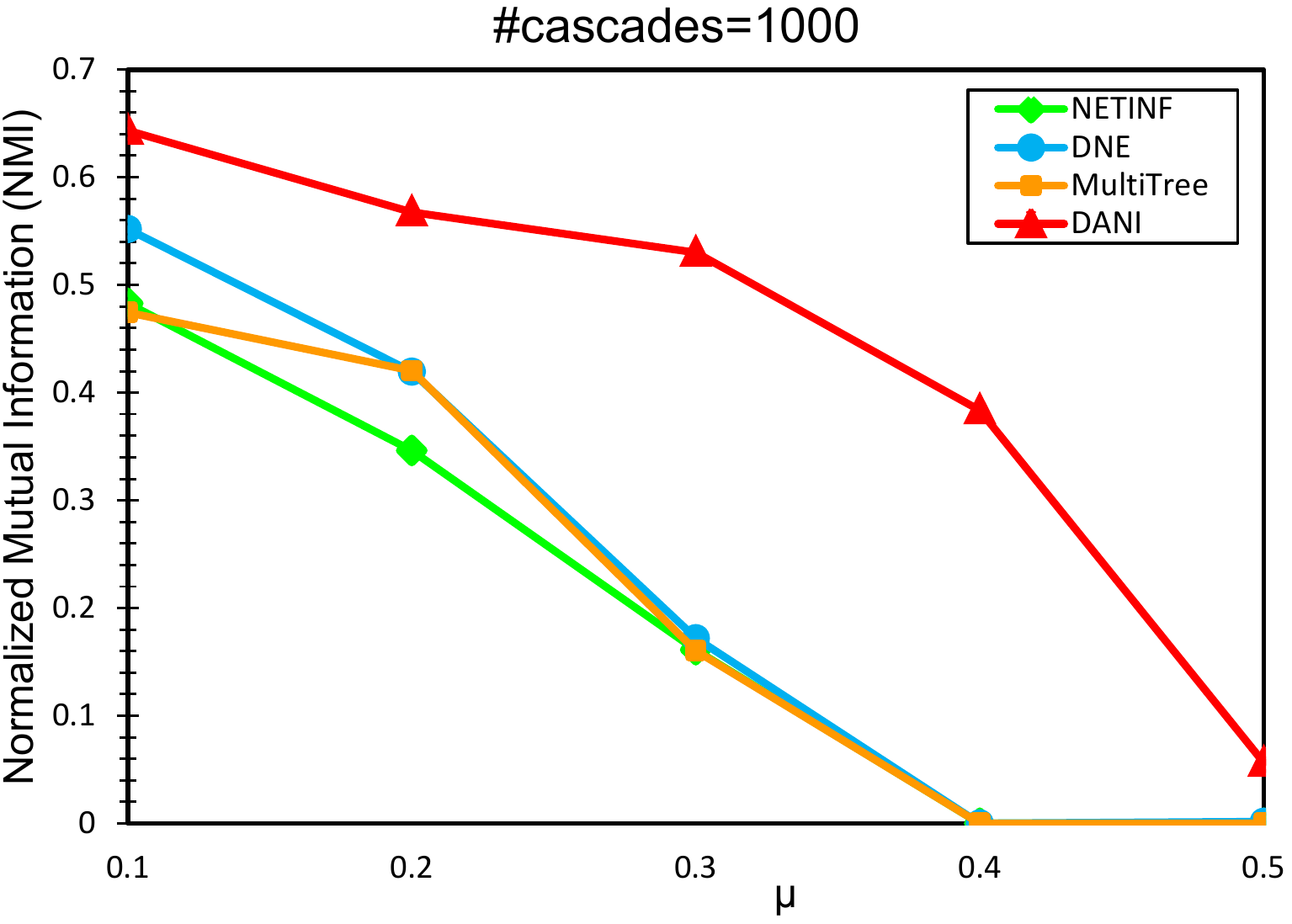}
\includegraphics[width=0.32\textwidth]{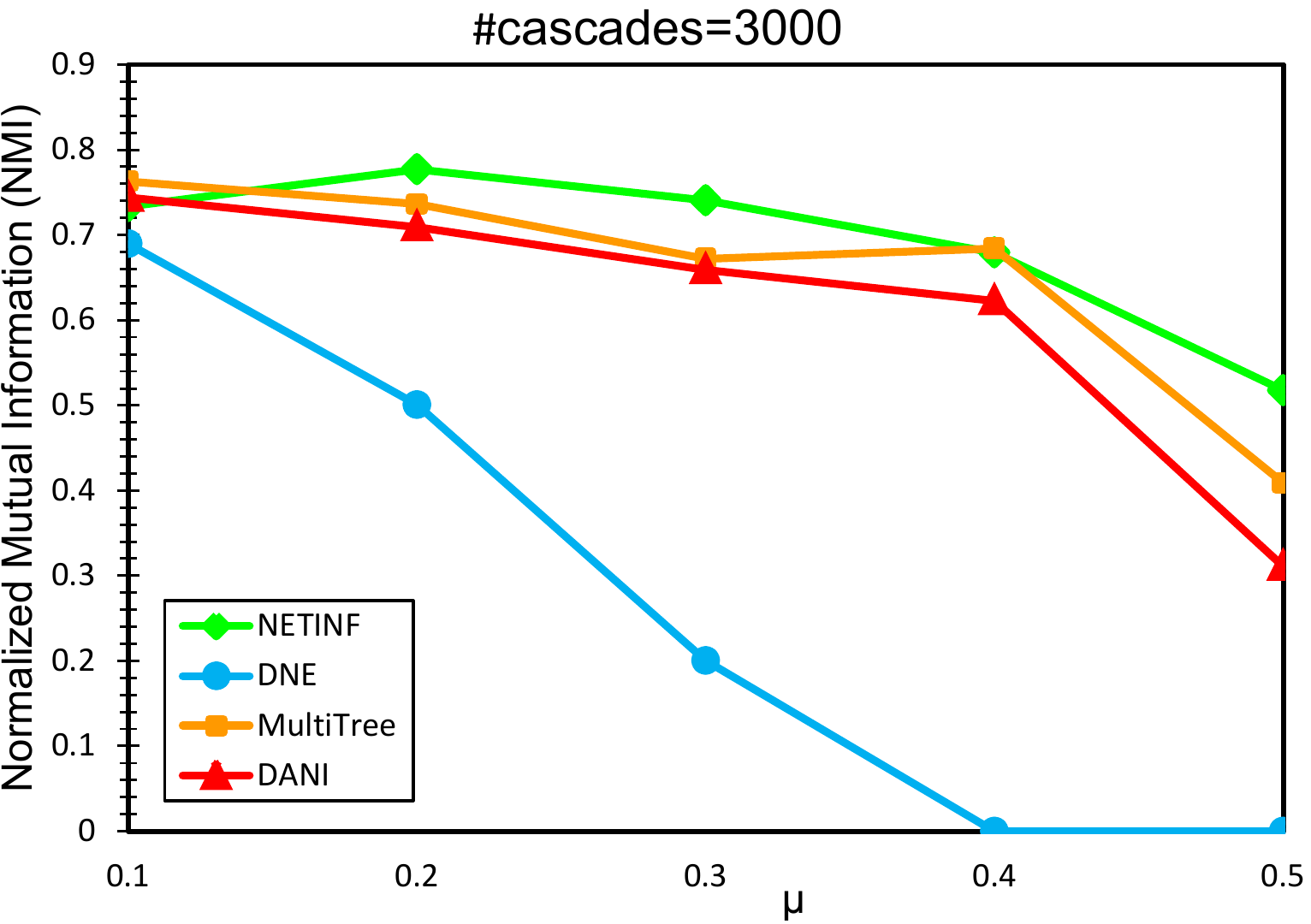}
\includegraphics[width=0.32\textwidth]{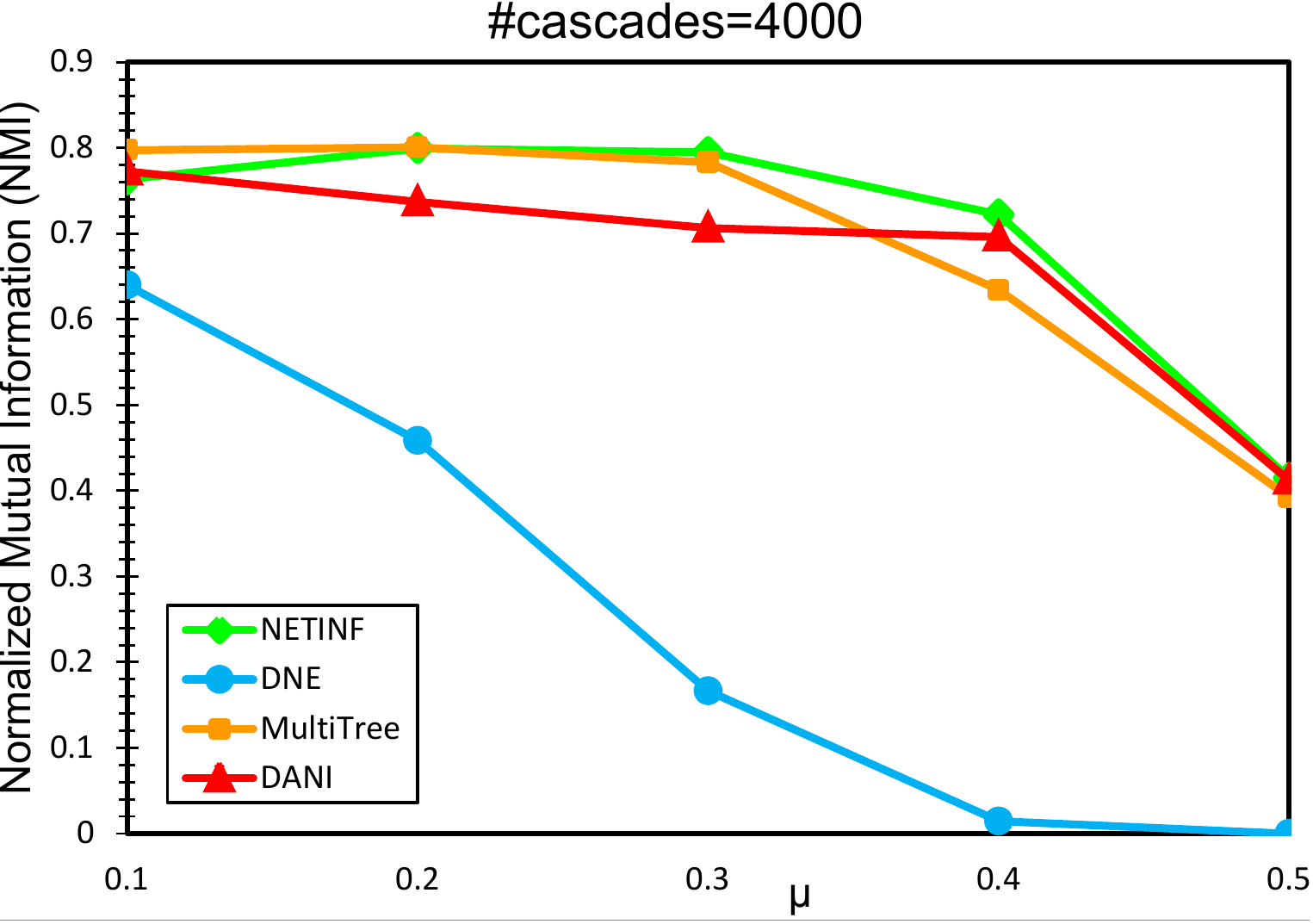}
\includegraphics[width=0.32\textwidth]{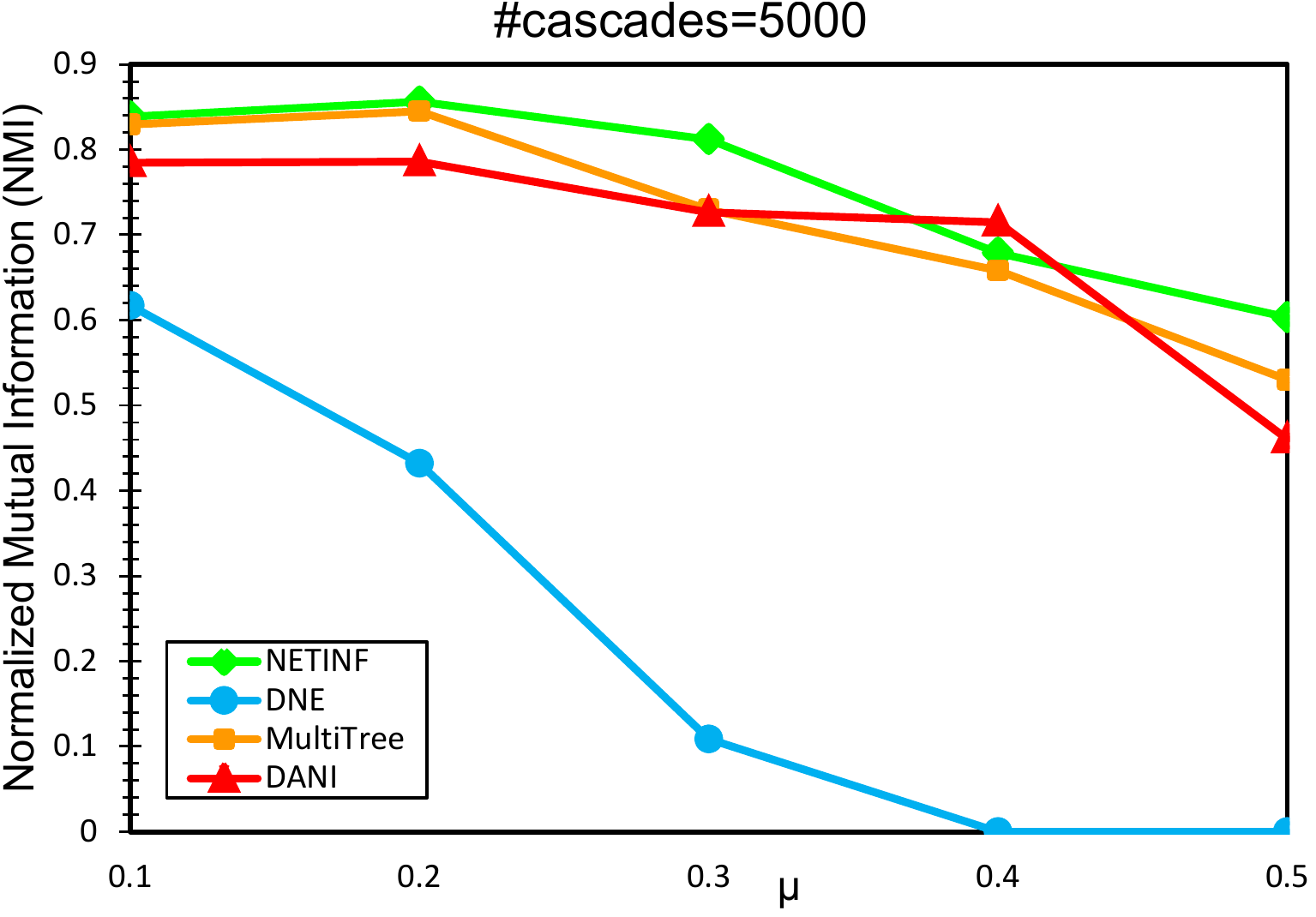}
\includegraphics[width=0.32\textwidth]{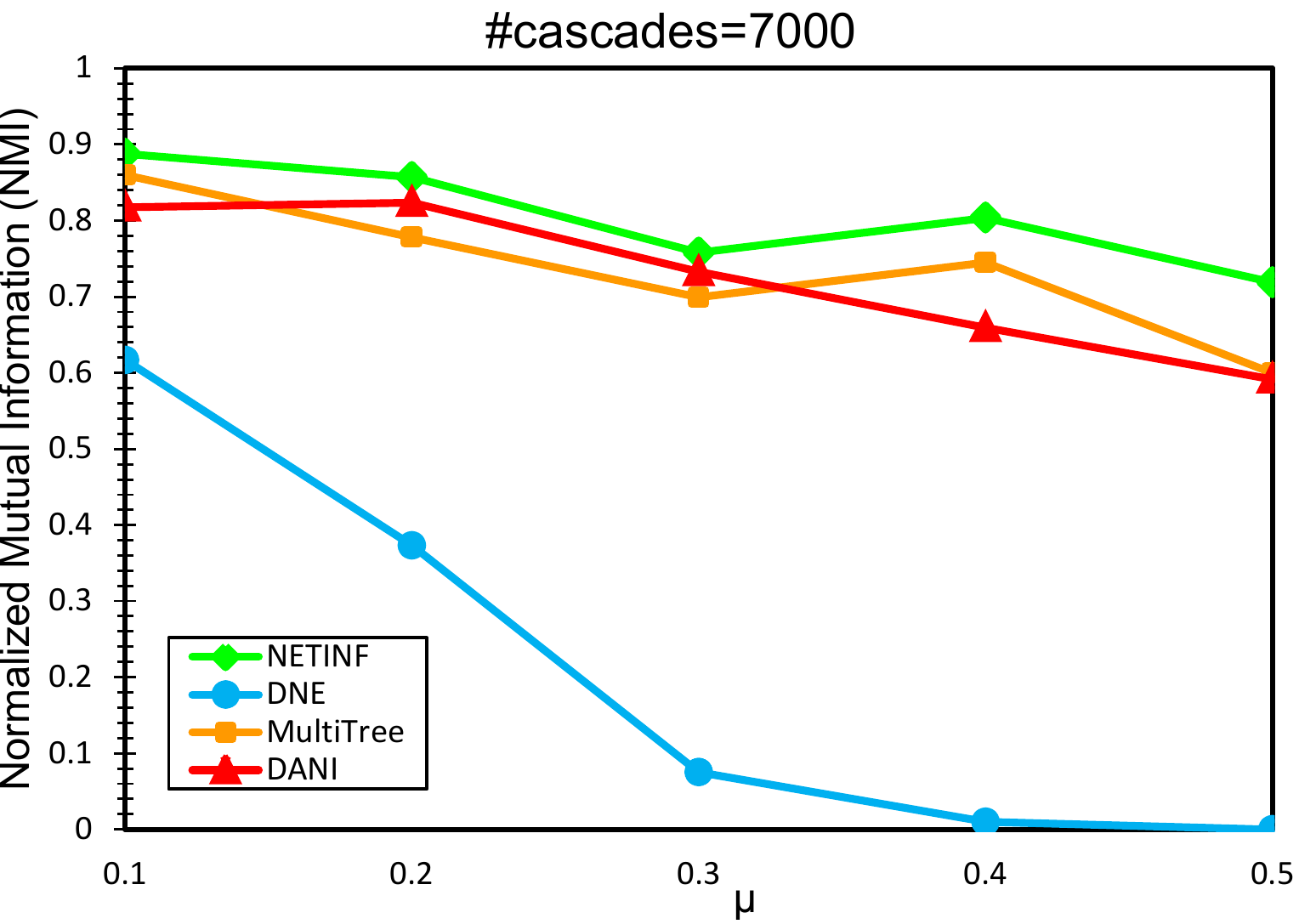}
\includegraphics[width=0.32\textwidth]{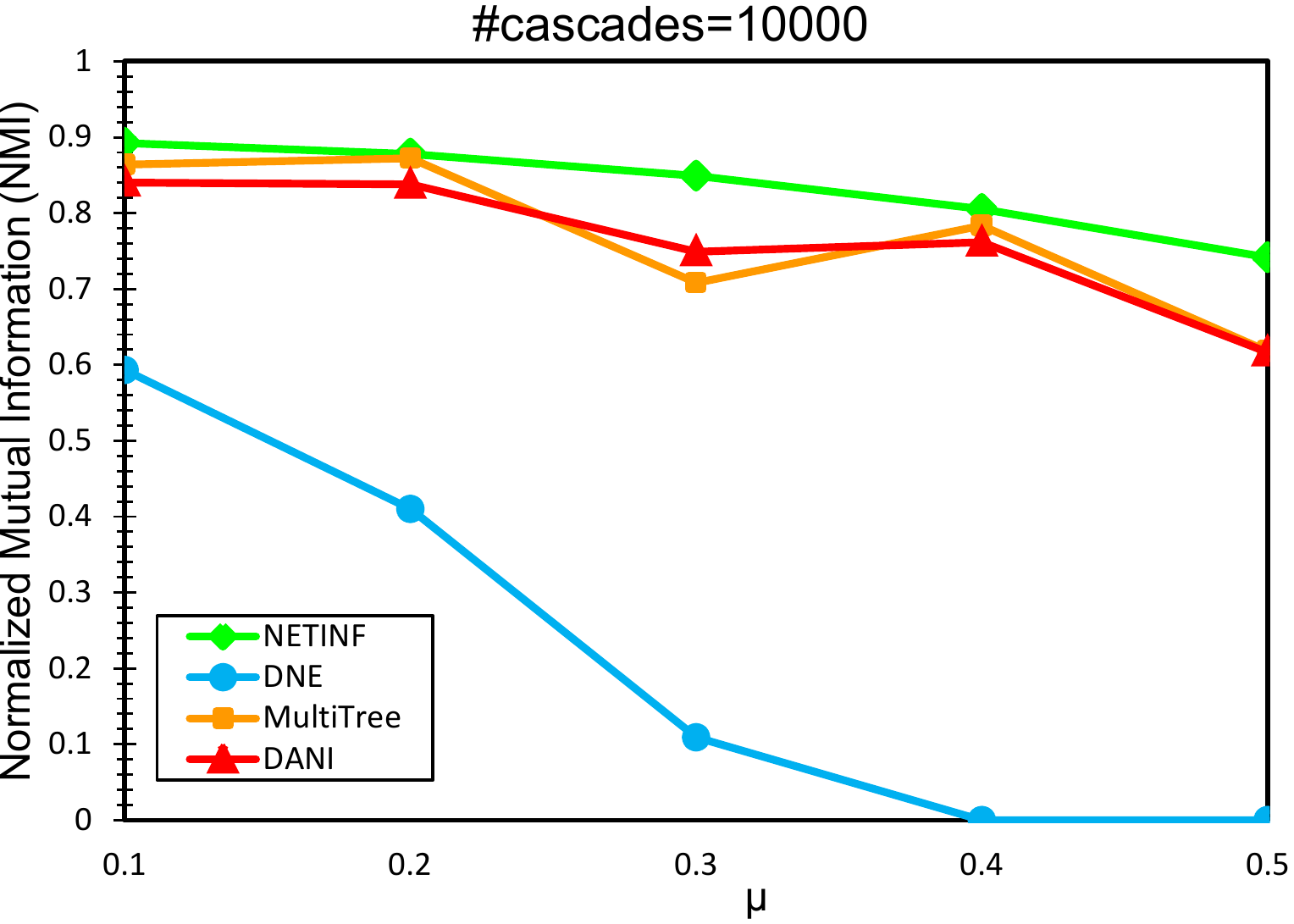}
\includegraphics[width=0.32\textwidth]{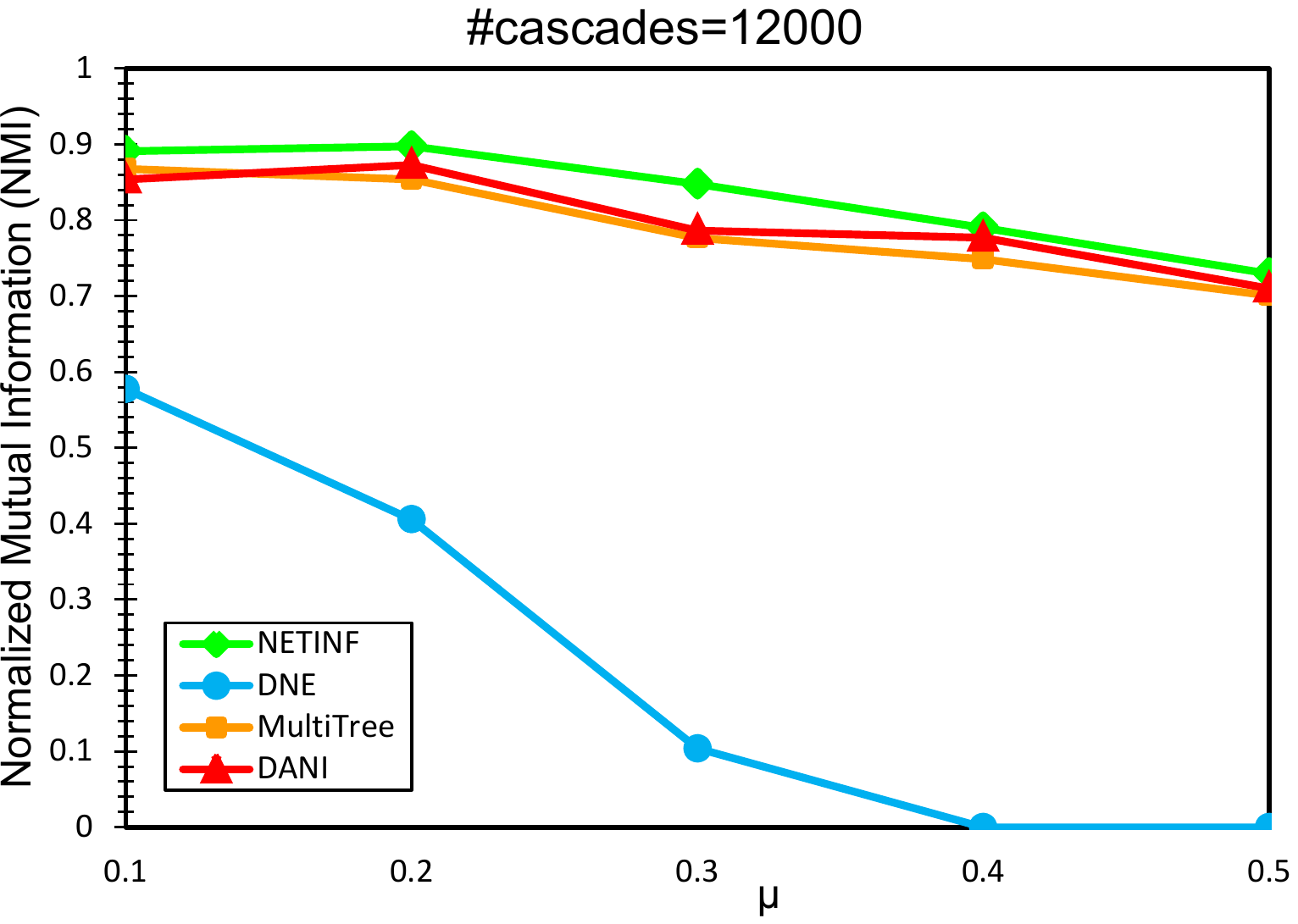}
\includegraphics[width=0.32\textwidth]{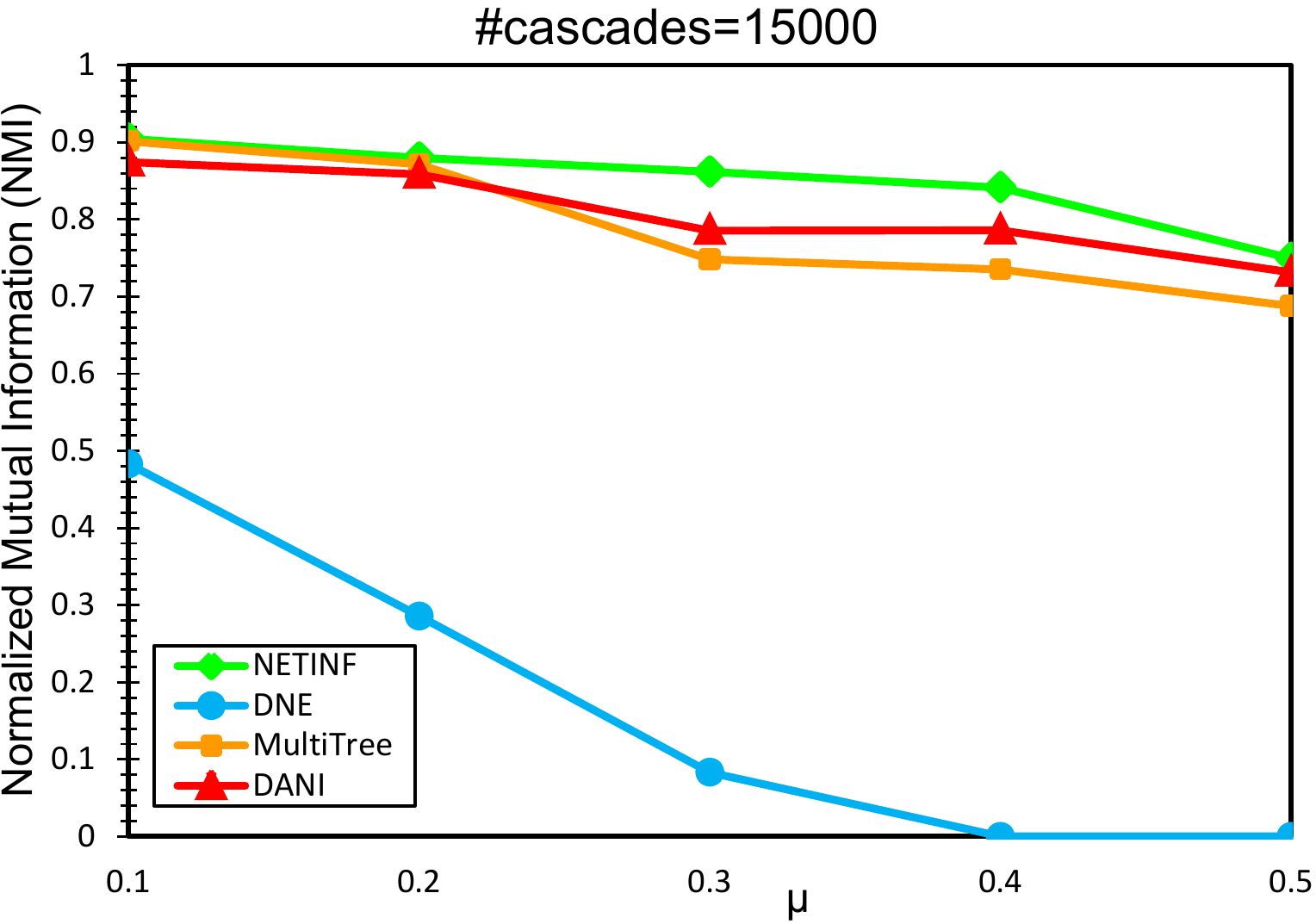}
\includegraphics[width=0.32\textwidth]{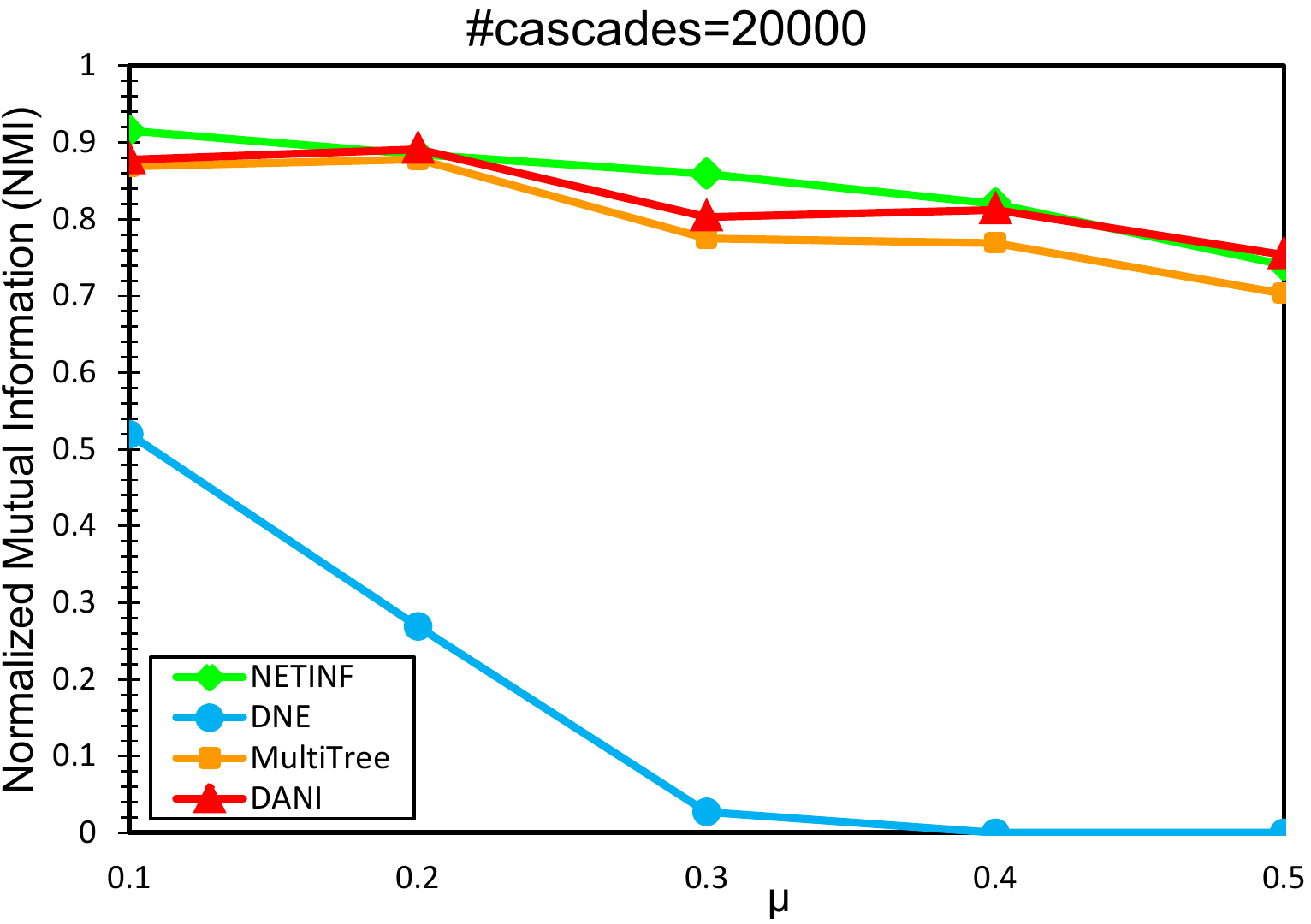}
\caption{\label{fig:NMILFR} (Color online) Normalized mutual information versus different community structures for various numbers of cascades in the LFR-benchmark. Figures show averages over 10 runs.}
\end{center}
\end{figure}

\paragraph{\textbf{NMI}:}
Fig.~\ref{fig:NMILFR} illustrates the values of NMI on each inferred LFR-benchmark network with varying values for $\mu$, and different number of cascades. As it can be observed, specially for fewer number of cascades, DANI performs perfectly. For higher number of cascades, the NMI values for DANI is closer to NMI of NETINF and MultiTree, while DNE has lower NMI values in all cases. Figs.~\ref{fig:LinkedInPlot} and ~\ref{fig:NewsoftheWorldPlot} show the NMI values of all algorithms for real datasets. 
As illustrated in Fig.~\ref{fig:NetScience}, for Netscience with artificial cascades, the running time of DANI is much lower, while its NMI is equal or more than NETINF.

For the LinkedIn dataset, as a real network, on average DANI performs $24.09\%$, $89.98\%$ and $23.89\%$ better than NETINF, MultiTree and DNE, respectively. Our improvements in News of the World dataset against NETINF, MultiTree, and DNE was $14.48\%$, $94.83\%$, and $15.84\%$, respectively. The differences between inferred and original networks, based on the criteria of Subsection \ref{sec:Evaluation metrics} is listed in Table~\ref{tab:Behavior}. As the results show, DANI has the closest values to the original network which indicates its capability in preserving the community structures.

\paragraph{\textbf{PWF}:}
For the NetScience dataset, DANI achieves higher PWF compared to the other methods at all number of cascades smaller than $5000$. For the LinkedIn network, on average DANI achieves $67.72\%$, $73.30\%$ and $29.08\%$ improvement in PWF against NETINF, MultiTree and DNE, respectively. Moreover, for News of the World, on average DANI achieves $31.29\%$, $35.57\%$ and $19.94\%$ improvement against NETINF, MultiTree and DNE, respectively. 

\paragraph{\textbf{Density}:}
The inferred network by DNE has a high density which is very different from the underlying network. However, NETINF, MultiTree and DANI achieve a density similar to the underlying network, while DANI performs better than the competing algorithms.

\paragraph{\textbf{Conductance}:}
The conductance of all competing algorithms are similar. Considering the average and variance of all intervals in one year, DANI performs marginally better than other methods for all datasets.

\paragraph{\textbf{Number of communities}:}
For this metric, DANI achieved the best performance in detecting the number of communities for the inferred network compared to the original network.

\begin{table}
\tbl{\label{tab:Behavior} Difference of metrics between communities at inferred and original networks for real datasets. Values are averaged over different number of cascades for the NetScience dataset, and over different months for LinkedIn and News of the World. Variances are shown in parentheses.}
{

\begin{tabular}[center]{|c|c|c|c|c|c|}
\hline

{Dataset}
&{Total Metrics} &{DNE}
&{NETINF}
&{DANI}
&{MultiTree}
\\
\hline
{NetScience}
&{\#Node}&453.77 (210502.3)&\textbf{445.41 (217805)}&\textbf{445.41 (217805)}&\textbf{445.41 (217805)}\\
\cline{2-6} 
&{Degree}&0.583 (0.0498)&0.575 (0.0645)&\textbf{0.529 (0.0714)}&0.569 (0.064)\\
\cline{2-6} 
&{Clusteringcoeff}&{0.457 (0.0395)}&{0.487 (0.0418)}&{0.359 (0.0733)}&{0.478 (0.0418)}\\
\cline{2-6} 
&{Density}&0.0403 (0.0004)&0.0401 (0.0003)&\textbf{0.0232 (0.0002)}&0.0433 (0.0006)\\
\cline{2-6} 
&{\#Community}&0.477 (0.060)&0.460 (0.076)&\textbf{0.404 (0.093)}&0.474 (0.069)\\
\cline{2-6} 
&{Conductance}&0.00515 (0.00002)&0.00703 (0.00004)&0.00786 (0.00006)&\textbf{0.00511 (0.00003)}\\
\hline

{LinkedIn}
&{\#Node}&109.73 (399.42)&53.64 (79.25)&\textbf{8.91 (19.70)}&54.18 (95.36)\\
\cline{2-6} 
&{Degree}&{0.747 (0.003)}&{0.807 (0.001)}&{\textbf{0.712 (0.002)}}&{0.805 (0.001)}\\
\cline{2-6} 
&{{Clusteringcoeff}}&{\textbf{0.0416 (0.0006)}}&{0.0757 (0.0018)}&{0.0643 (0.0004)}&{0.0772 (0.0018)}\\
\cline{2-6}  
&{Density}&0.24 (0.01)&0.10 (0.006)&\textbf{0.05 (0.002)}&0.09 (0.003)\\
\cline{2-6} 
&{\#Community}&\textbf{0.21 (0.02)}&0.45 (0.30)&0.30 (0.08)&0.38 (0.03)\\
\cline{2-6} 
&{Conductance}&0.04 (0.0007)&{0.03 (0.0003)}&\textbf{0.02 (0.0004)}&\textbf{0.02 (0.0003)}\\
\hline

{News of the World}
&{\#Node}&67.73 (2024.22)&25.64 (305.25)&\textbf{3.91 (11.49)}&26.00 (302)\\
\cline{2-6} 
&{{Degree}}&{0.888 (0.024)}&{0.925 (0.016)}&{\textbf{0.688 (0.007)}}&{0.918 (0.015)}\\
\cline{2-6} 
&{{Clusteringcoeff}}&{0.12 (0.004)}&{0.09 (0.004)}&{\textbf{0.09 (0.001)}}&{0.09 (0.004)}\\
\cline{2-6}  
&{Density}&0.315 (0.0193)&0.124 (0.005)&\textbf{0.107 (0.003)}&0.134 (0.007)\\
\cline{2-6} 
&{\#Community}&0.17 (0.01)&0.29 (0.02)&\textbf{0.16 (0.01)}&0.30 (0.03)\\
\cline{2-6} 
&{Conductance}&0.035 (0.0006)&0.029 (0.0007)&\textbf{0.029 (0.0006)}&0.038 (0.0012)\\
\hline
\end{tabular}
}
\end{table}

\section{CONCLUSION}
\label{sec:conclusion}

Many works have been done to infer networks by using information diffusion, but they often neglect preserving the features of underlying networks. Community structure is one of the most important features in social networks and community detection algorithms need the network topology to extract their communities. Therefore, these algorithms cannot be applied on networks derived by the previous proposed network inference methods. 

In this paper, we introduced a new algorithm to infer networks which maintains the community structure. To verify this claim, we applied the proposed (DANI), NETINF, DNE, and MultiTree methods to real and artificial networks, and compared the community structure of original and inferred networks with each other. Moreover, we utilized some of the main community structure metrics to evaluate the inferred networks by different methods. The results showed that our algorithm accurately infers networks while preserving the community structures. In addition, DANI consumed less time to achieve the same or higher accuracy than the competing methods. Another advantage of the proposed method is the fact that its performance is independent of the length of cascades.

As a future work, we are trying to detect communities just based on diffusion information without the need for inferring networks. In other words, we try to find the community of each node only based on the diffusion process.

\bibliographystyle{ACM-Reference-Format-Journals}
\bibliography{acmsmall-sample-bibfile}



\medskip

\end{document}